\documentclass[ALICE,manyauthors]{cernphprep}
\usepackage[comma,square,numbers,sort&compress]{natbib}
\usepackage{hyperref}
\usepackage{lineno}
\usepackage{xspace}
\usepackage{xcolor}
\usepackage[T1]{fontenc}
\begin{document}
%

\newcommand{\pp}           {pp\xspace}
\newcommand{\ppbar}        {\mbox{$\mathrm {p\overline{p}}$}\xspace}
\newcommand{\ccbar}        {\mbox{$\mathrm {c\overline{c}}$}\xspace}
\newcommand{\bbbar}        {\mbox{$\mathrm {b\overline{b}}$}\xspace}
\newcommand{\XeXe}         {\mbox{Xe--Xe}\xspace}
\newcommand{\PbPb}         {\mbox{Pb--Pb}\xspace}
\newcommand{\pA}           {\mbox{pA}\xspace}
\newcommand{\pPb}          {\mbox{p--Pb}\xspace}
\newcommand{\AuAu}         {\mbox{Au--Au}\xspace}
\newcommand{\dAu}          {\mbox{d--Au}\xspace}

\newcommand{\s}            {\ensuremath{\sqrt{s}}\xspace}
\newcommand{\snn}          {\ensuremath{\sqrt{s_{\mathrm{NN}}}}\xspace}
\newcommand{\pt}           {\ensuremath{p_{\rm T}}\xspace}
\newcommand{\meanpt}       {$\langle p_{\mathrm{T}}\rangle$\xspace}
\newcommand{\ycms}         {\ensuremath{y_{\rm CMS}}\xspace}
\newcommand{\ylab}         {\ensuremath{y_{\rm lab}}\xspace}
\newcommand{\etarange}[1]  {\mbox{$\left | \eta \right |~<~#1$}}
\newcommand{\yrange}[1]    {\mbox{$\left | y \right |~<~#1$}}
\newcommand{\dndy}         {\ensuremath{\mathrm{d}N_\mathrm{ch}/\mathrm{d}y}\xspace}
\newcommand{\dndeta}       {\ensuremath{\mathrm{d}N_\mathrm{ch}/\mathrm{d}\eta}\xspace}
\newcommand{\avdndeta}     {\ensuremath{\langle\dndeta\rangle}\xspace}
\newcommand{\dNdy}         {\ensuremath{\mathrm{d}N_\mathrm{ch}/\mathrm{d}y}\xspace}
\newcommand{\dndydpt}
{\ensuremath{\mathrm{d}^2\sigma_{\jpsi}/\mathrm{d}\pt \mathrm{d}y}\xspace}
\newcommand{\Npart}        {\ensuremath{N_\mathrm{part}}\xspace}
\newcommand{\Ncoll}        {\ensuremath{N_\mathrm{coll}}\xspace}
\newcommand{\dEdx}         {\ensuremath{\textrm{d}E/\textrm{d}x}\xspace}
\newcommand{\RpPb}         {\ensuremath{R_{\rm pPb}}\xspace}
\newcommand{\dsdy}         {\ensuremath{\textrm{d}\sigma/\textrm{d}y}\xspace}
\newcommand{\fb}           {\ensuremath{f_{\rm B}}}
\newcommand{\fbp}           {\ensuremath{f'_{\rm B}}}
\newcommand{\mee}          {\ensuremath{m_{\rm ee}}}
\newcommand{\x}            {\ensuremath{x}}
\newcommand{\fsig}         {\ensuremath{f_{\rm Sig}}}
\newcommand{\hb}           {\ensuremath{h_{\rm B}}}
\newcommand{\question}[1]     {\textcolor{blue}{#1}}
\newcommand{\update}[1]     {\textcolor{red}{#1}}

\newcommand{\nineH}        {$\sqrt{s}~=~0.9$~Te\kern-.1emV\xspace}
\newcommand{\seven}        {$\sqrt{s}~=~7$~Te\kern-.1emV\xspace}
\newcommand{\twoH}         {$\sqrt{s}~=~0.2$~Te\kern-.1emV\xspace}
\newcommand{\twosevensix}  {$\sqrt{s}~=~2.76$~Te\kern-.1emV\xspace}
\newcommand{\five}         {$\sqrt{s}~=~5.02$~Te\kern-.1emV\xspace}
\newcommand{\twosevensixnn}{$\sqrt{s_{\mathrm{NN}}}~=~2.76$~Te\kern-.1emV\xspace}
\newcommand{\fivenn}       {$\sqrt{s_{\mathrm{NN}}}~=~5.02$~Te\kern-.1emV\xspace}
\newcommand{\LT}           {L{\'e}vy-Tsallis\xspace}
\newcommand{\GeVc}         {Ge\kern-.1emV/$c$\xspace}
\newcommand{\MeVc}         {Me\kern-.1emV/$c$\xspace}
\newcommand{\TeV}          {Te\kern-.1emV\xspace}
\newcommand{\GeV}          {Ge\kern-.1emV\xspace}
\newcommand{\MeV}          {Me\kern-.1emV\xspace}
\newcommand{\GeVmass}      {Ge\kern-.2emV/$c^2$\xspace}
\newcommand{\MeVmass}      {Me\kern-.2emV/$c^2$\xspace}
\newcommand{\lumi}         {\ensuremath{\mathcal{L}}\xspace}

\newcommand{\ITS}          {\rm{ITS}\xspace}
\newcommand{\TOF}          {\rm{TOF}\xspace}
\newcommand{\ZDC}          {\rm{ZDC}\xspace}
\newcommand{\ZDCs}         {\rm{ZDCs}\xspace}
\newcommand{\ZNA}          {\rm{ZNA}\xspace}
\newcommand{\ZNC}          {\rm{ZNC}\xspace}
\newcommand{\SPD}          {\rm{SPD}\xspace}
\newcommand{\SDD}          {\rm{SDD}\xspace}
\newcommand{\SSD}          {\rm{SSD}\xspace}
\newcommand{\TPC}          {\rm{TPC}\xspace}
\newcommand{\TRD}          {\rm{TRD}\xspace}
\newcommand{\VZERO}        {\rm{V0}\xspace}
\newcommand{\VZEROA}       {\rm{V0A}\xspace}
\newcommand{\VZEROC}       {\rm{V0C}\xspace}
\newcommand{\Vdecay} 	   {\ensuremath{V^{0}}\xspace}

\newcommand{\ee}           {\ensuremath{\text{e}^{+}\text{e}^{-}}\xspace}
\newcommand{\pip}          {\ensuremath{\pi^{+}}\xspace}
\newcommand{\pim}          {\ensuremath{\pi^{-}}\xspace}
\newcommand{\kap}          {\ensuremath{\rm{K}^{+}}\xspace}
\newcommand{\kam}          {\ensuremath{\rm{K}^{-}}\xspace}
\newcommand{\pbar}         {\ensuremath{\rm\overline{p}}\xspace}
\newcommand{\kzero}        {\ensuremath{{\rm K}^{0}_{\rm{S}}}\xspace}
\newcommand{\lmb}          {\ensuremath{\Lambda}\xspace}
\newcommand{\almb}         {\ensuremath{\overline{\Lambda}}\xspace}
\newcommand{\Om}           {\ensuremath{\Omega^-}\xspace}
\newcommand{\Mo}           {\ensuremath{\overline{\Omega}^+}\xspace}
\newcommand{\X}            {\ensuremath{\Xi^-}\xspace}
\newcommand{\Ix}           {\ensuremath{\overline{\Xi}^+}\xspace}
\newcommand{\Xis}          {\ensuremath{\Xi^{\pm}}\xspace}
\newcommand{\Oms}          {\ensuremath{\Omega^{\pm}}\xspace}
\newcommand{\degree}       {\ensuremath{^{\rm o}}\xspace}
\newcommand{\jpsi}        {\ensuremath{\rm J/\psi}\xspace}
\newcommand{\pJPsi}        {\ensuremath{\rm J/\psi}\xspace}
\newcommand{\DMeson}       {\ensuremath{\rm D}\xspace}

\begin{titlepage}
\PHyear{2021}       
\PHnumber{152}      
\PHdate{26 July}  


\title{Prompt and non-prompt \pJPsi\ production cross sections at midrapidity in proton\textendash proton collisions at \s = 5.02 and 13 TeV}
\ShortTitle{Non-prompt \pJPsi\ production in proton\textendash proton collisions}   

\Collaboration{ALICE Collaboration\thanks{See Appendix~\ref{app:collab} for the list of collaboration members}}
\ShortAuthor{ALICE Collaboration} 

\begin{abstract}

	The production of \pJPsi\ is measured at midrapidity ($|y|<0.9$) in proton\textendash proton collisions at \mbox{\s = 5.02} and 13 TeV, through the dielectron decay channel, using the ALICE detector at the Large Hadron Collider. 
	The data sets used for the analyses correspond to integrated luminosities of  \mbox{$\mathcal{L}_{\rm int}$ = 19.4~$\pm$~0.4~nb$^{-1}$} and \mbox{ $\mathcal{L}_{\rm int}$ = 32.2~$\pm$~0.5~nb$^{-1}$} at \s~=~5.02 and 13~TeV, respectively. 
	The fraction of non-prompt \pJPsi\ mesons, i.e.\ those originating from the decay of beauty hadrons, is measured down to a transverse momentum \pt = 2~\GeVc\ (1~\GeVc) at  \s~=~5.02 TeV (13 TeV). The \pt\ and rapidity ($y$) differential cross sections, as well as the corresponding values integrated over \pt and $y$, are carried out separately for prompt and non-prompt \pJPsi\ mesons. 
The results are compared with measurements from other experiments and theoretical calculations based on quantum chromodynamics (QCD). The shapes of the \pt\ and $y$ distributions of beauty quarks predicted by state-of-the-art perturbative QCD models are used to extrapolate an estimate of the \bbbar pair cross section at midrapidity and in the total phase space. 
The total \bbbar cross sections are found to be 
$\sigma_{\rm b \overline{\rm b}} = 541 \pm 45 (\rm stat.) \pm 69 (\rm syst.)_{-12}^{+10} (\rm extr.)~{\rm \mu b}$ and 
	$\sigma_{\rm b \overline{\rm b}}~=~218 \pm 37 (\rm stat.) \pm 31 (\rm syst.)_{-9.1}^{+8.2} (\rm extr.)~{\rm \mu b}$ 
	at \s = 13 and 5.02 TeV, respectively.  
	The value obtained from the combination of ALICE and LHCb measurements in pp collisions at \s = 13 TeV is also provided.
\end{abstract}
\end{titlepage}

\setcounter{page}{2} 

\section{Introduction}
\label{Sec:Intro} 

The study of the production of hidden and open heavy-flavour hadrons in proton\textendash proton (pp) collisions provides an essential test of quantum chromodynamics (QCD), involving both the perturbative and non-perturbative regimes of this theory. Experimentally, the reconstruction of the lightest charmonium vector state, the \pJPsi meson, produced in pp collisions at the energies of the Large Hadron Collider (LHC) gives access to both the physics of charmonium systems and that of beauty-quark production. 
Indeed, direct \pJPsi mesons and feed-down from higher mass charmonium states such as $\chi_{\rm c}$ and $\psi(\rm 2S)$, which are denoted as the ``prompt'' component, can be experimentally separated from the contribution from long-lived weak decays of beauty hadrons, denoted as the ``non-prompt'' component. In addition, due to the large rest mass of the \pJPsi as compared to the other beauty-hadron decay products, the \pJPsi momentum vector is very close to those of the decaying beauty-hadron, making the non-prompt \pJPsi measurement a good tool to study the production of beauty-flavour hadrons~\cite{PhysRevD.71.032001}.

Due to the very different energy and time scales involved in prompt charmonium production~\cite{Ellis:318585}, phenomenological models assume that the cross section factorises into a hard term, describing the initial production of the \ccbar pair, and a soft term accounting for the subsequent evolution into a bound state. While the production of \ccbar pairs can be computed within perturbative QCD, their evolution to a bound state involves long-distance physics which is non-perturbative and relies largely on fits to experimental measurements. A detailed overview of this field of study can be found in Refs.~\cite{Brambilla:2010cs,Brambilla:2014jmp,Andronic:2015wma,Chen:2021tmf}. There are a few different approaches employed for the description of quarkonium production, namely the Colour Singlet Model (CSM)~\cite{Baier:1981uk}, the Colour Evaporation Model (CEM)~\cite{FRITZSCH1977217,AMUNDSON1997323}, and the Non-Relativistic QCD model (NRQCD)~\cite{PhysRevD.55.5853}. 
High precision measurements, in particular at the LHC~\cite{Aaij2013,BAbelev2012,Acharya:2019lkw,Adam:2016rdg,Acharya:2017hjh,Aaboud:2017cif,Sirunyan:2017mzd,Sirunyan:2017isk,ALICE:2021qlw,LHCb:2021pyk,Abelev:2012gx,Abelev:2014qha,Aaij:2011jh,Khachatryan:2010yr,Aad:2011sp,CMS:2015lbl,Adam:2015rta,Aaij:2013yaa,ATLAS:2015zdw,Alice13TeV,Aaij:2015rla,Sirunyan:2017qdw,PhysRevLett.108.082001,Acharya:2018uww,Chatrchyan:2013cla,Aaij:2013nlm}, enabled significant improvements of the theoretical description of charmonium production in all of these approaches as shown for CEM in Refs.~\cite{Ma:2016exq,PhysRevD.98.114029}, for CSM in Refs.~\cite{Artoisenet:2007xi,Campbell:2007ws,Artoisenet:2008fc} and NRQCD in Refs.~\cite{Ma:2010jj,Butenschoen:2010rq,Ma:2010yw,Chao:2012iv,Butenschoen:2012px,Butenschoen:2012qr,Gong:2012ug,Butenschoen:2011yh,Bodwin:2014gia,Faccioli:2014cqa,Bodwin:2015iua,Faccioli:2017hym,Faccioli:2018uik}.
However, a full description of the data, in particular down to zero transverse momentum is still difficult~\cite{Acharya:2018uww,Alice13TeV,ALICE:2021qlw}. 
Within the NRQCD factorisation formalism, the long-distance effects leading to the charmonium hadronisation are implemented using the so called long distance matrix elements (LDMEs) which are fixed in fits to experimental data and are hypothesized to be universal.
Several global fit analyses have been performed in the past decade using NLO short distance coefficients (SDCs), providing a simultaneous description of cross sections and polarisations measured at the LHC, with a good fit quality, as long as the fits are restricted to the region of $\pt>8$~GeV/$c$~\cite{Ma:2010jj,Butenschoen:2010rq,Ma:2010yw,Chao:2012iv,Butenschoen:2012px,Butenschoen:2012qr,Gong:2012ug,Butenschoen:2011yh,Bodwin:2014gia,Faccioli:2014cqa,Bodwin:2015iua,Faccioli:2017hym,Faccioli:2018uik}.
The production at low transverse momentum, which dominates the total cross section, requires a special treatment since the 
collinear factorisation approach~\cite{Collins:1989gx} is no longer applicable. Combining the Color Glass Condensate (CGC) effective field theory at small $x$~\cite{Gelis:2010nm} and NRQCD formalisms, a new factorisation framework for quarkonium production was proposed, which provides a  satisfactory description of charmonia at low \pt~\cite{Ma:2018qvc,Chen:2021tmf}.

The inclusive production of open heavy-flavour hadrons in hadronic collisions is computed using the collinear factorisation approach~\cite{Collins:1989gx} as a convolution
of the parton distribution functions of the incoming hadrons, the hard parton\textendash parton scattering cross section computed perturbatively, and the fragmentation process describing the non-perturbative evolution of a charm- or beauty-quark into an open heavy-flavour hadron.
These calculations are implemented at the next-to-leading order (NLO) accuracy in the general-mass variable-flavour-number scheme
(GM-VFNS)~\cite{Benzke:2019usl,Kniehl:2005mk}, and at NLO with an all-order resummation to next-to-leading log (NLL) accuracy in the limit where the \pt of the heavy quark is much larger than its mass in the FONLL resummation approach~\cite{Cacciari_2001,Cacciari:2012ny}.
Recent calculations with next-to-next-to-leading-order (NNLO) QCD radiative corrections are implemented for the beauty-quark production cross section~\cite{Catani:2020kkl}.
Other predictions are also performed in the leading order (LO) approximation through the $k_{\rm T}$-factorisation framework~\cite{Catani:1990eg}.
All these computations describe, albeit within large theoretical uncertainties, the production cross sections of open heavy-flavour hadrons measured in pp and p$\overline{\rm p}$ collisions
in different kinematic domains at centre-of-mass energies ranging from 0.2 to 13 TeV~\cite{Andronic:2015wma,PhysRevLett.119.169901,Khachatryan:2016csy}. 
Non-prompt \pJPsi production is directly related to open beauty-hadron production, and can be used to estimate the latter after an extrapolation.
The measurements of the total beauty-quark production cross sections  are less sensitive to the non-perturbative hadronisation effects than the total charm\textendash quark production, which makes them a good test for QCD in the perturbative regime.
In the context of the LHC heavy-ion physics programme, the measurement of beauty production in pp collisions is crucial for studying  both cold and hot nuclear matter effects, as they provide a reference for the beauty-hadron production measurements in proton\textendash nucleus and nucleus\textendash nucleus collisions. 

Before the start of the LHC, \pJPsi production was extensively studied in \ppbar and \pp collisions at the Tevatron~\cite{PhysRevD.71.032001,PhysRevLett.99.132001,Abachi:1996jq,PhysRevLett.82.35} and at RHIC~\cite{PhysRevLett.98.232002,Tang:2020ame}. At the LHC, the \pJPsi\ transverse momentum and rapidity differential 
production cross sections have been measured in \pp collisions at several centre-of-mass energies, namely \s = 2.76~TeV~\cite{Aaij2013,BAbelev2012}, 5.02~TeV~\cite{Acharya:2019lkw,Adam:2016rdg,Acharya:2017hjh,Aaboud:2017cif,Sirunyan:2017mzd,Sirunyan:2017isk,ALICE:2021qlw,LHCb:2021pyk}, 7~TeV~\cite{Abelev:2012gx,Abelev:2014qha,Aaij:2011jh,Khachatryan:2010yr,Aad:2011sp,ATLAS:2015zdw,CMS:2015lbl}, 8~TeV~\cite{Adam:2015rta,Aaij:2013yaa,ATLAS:2015zdw}, and 13~TeV~\cite{Alice13TeV,Acharya:2017hjh,Aaij:2015rla,Sirunyan:2017qdw}. Experimental measurements were also extended to other observables, such as polarisation which was measured by ALICE~\cite{PhysRevLett.108.082001,Acharya:2018uww}, CMS~\cite{Chatrchyan:2013cla}, and LHCb~\cite{Aaij:2013nlm} in pp collisions at \s = 7 and 8 TeV. 
At the centre-of-mass energies discussed in this article, the prompt and non-prompt components of the \pJPsi\ production cross section at midrapidity were previously studied in pp collisions at \s = 5.02 TeV by the ATLAS~\cite{Aaboud:2017cif} and CMS~\cite{Sirunyan:2017mzd,Sirunyan:2017isk} collaborations and at \s = 13 TeV by CMS~\cite{Sirunyan:2017qdw}. At forward rapidity ($2 < y < 4.5$), the LHCb collaboration reported prompt and non-prompt \pJPsi\ measurements at \s = 13 TeV~\cite{Aaij:2015rla} and at \s = 5 TeV~\cite{LHCb:2021pyk}.

In this article, the prompt and non-prompt \pJPsi\ cross section measurements performed at midrapidity ($|y| < 0.9$) at \s = 5.02 and 13 TeV via the dielectron decay channel are reported. Measurements are carried out down to a transverse momentum of 2 GeV/$c$ at \s = 5.02 TeV and 1 GeV/$c$ at 13 TeV.
They are complementary to the existing ATLAS and CMS measurements available for \pt\ $>$ 8 GeV/$c$ and 6.5 GeV/$c$, respectively. 
The low-\pt reach for non-prompt \pJPsi\ allows the derivation of the ${\rm d}\sigma_{\rm b \overline{\rm b}}/{\rm d}y$ at midrapidity and of the total \bbbar cross section at both energies \s = 5.02 and 13 TeV. 

The article is organised as follows: the ALICE apparatus and data samples are described in Section~\ref{Sec:DataAndApp},
the data analysis is detailed in Section~\ref{Sec:dataAnalysis}, results are discussed in Section~\ref{Sec:results} and compared to
existing measurements and to theoretical model calculations, and finally in Section~\ref{Sec:summary} conclusions are drawn.

\section{Apparatus and data samples}
\label{Sec:DataAndApp}

The ALICE apparatus comprises a central barrel placed in a solenoidal magnet that generates a constant field of $B$~=~0.5~T oriented along the beam axis ($z$), a muon spectrometer at forward rapidity, and a set of forward and backward detectors used for triggering and event characterisation. A detailed description of the apparatus and its performance can be found in Refs.~\cite{Collaboration_2008,Abelev:2014ffa}. The main detectors of the central barrel employed for the reconstruction of the \pJPsi via the \ee decay channel are the Inner Tracking System (ITS)~\cite{Aamodt:2010aa} and the Time Projection Chamber (TPC)~\cite{ALME2010316}. Both are used for track reconstruction, while the TPC is also used for electron identification and the ITS for primary and secondary vertex reconstruction. The ITS is composed of six cylindrical layers of
high-resolution silicon tracking detectors. The innermost layers consist of two arrays of hybrid Silicon Pixel Detectors (SPD), located at an average radial distance $r$ of 3.9 and 7.6 cm from the beam axis and covering the pseudorapidity intervals $|\eta| < 2.0$ and $|\eta| < 1.4$, respectively. The SPD provides the spatial resolution to separate on a statistical basis the prompt and non-prompt \pJPsi components. 
The outer layers of the ITS are composed of Silicon Drift Detector (SDD) and Silicon Strip Detector (SSD), with the outermost layer having a radius $r$ = 43 cm. 
The TPC is a large cylindrical drift detector with radial and longitudinal sizes of about $85 < r < 250$ cm and $-250 < z < 250$ cm, respectively. It is the main tracking device and its readout is segmented radially in pad rows, providing up to 159 space points per charged-particle track. The identification of
charged tracks is performed via the measurement of the specific ionisation energy loss \dEdx in the TPC gas. 

The events are selected using a minimum bias trigger
provided by the V0 detectors~\cite{Abbas:2013taa}, defined as the coincidence in signals between its two subsystems, V0C and V0A. The two V0 subsystems are scintillator arrays placed on both sides of the nominal interaction point at $z$ = $-$90 and +340 cm, covering the ranges $-3.7 < \eta < -1.7$ and $2.8 < \eta < 5.1$, respectively. The settings of the minimum bias trigger, which is fully efficient in inelastic collisions producing a \pJPsi, are identical at the two centre-of-mass energies.
The results in pp collisions at \s = 5.02 TeV are obtained using data recorded by ALICE 
in 2017, whereas the measurements carried out at \s = 13 TeV  are based on data samples collected during the years 2016$-$2018. The event samples, which are the same as those used for the published inclusive \pJPsi analyses at both energies~\cite{Acharya:2019lkw,Alice13TeV}, correspond to integrated luminosities of $\mathcal{L}_{\rm int}$ = 19.4 $\pm$ 0.4 nb$^{-1}$~\cite{ALICE-PUBLIC-2018-014} and $\mathcal{L}_{\rm int}$ = 32.2 $\pm$ 0.5 nb$^{-1}$~\cite{ALICE-PUBLIC-2021-005} at \s = 5.02 and 13 TeV, respectively.

\section{Data analysis}
\label{Sec:dataAnalysis}

Event selection and track quality requirements used in these analyses are identical to those used for the corresponding inclusive \pJPsi\ cross section analyses at \s\ = 5.02~\cite{Acharya:2019lkw} and 13 TeV~\cite{Alice13TeV}. In particular, the events, besides fulfilling the minimum bias trigger condition, 
are selected offline by requesting the collision vertex to be within the longitudinal interval $|z_{\rm vtx}| < 10$ cm around the nominal interaction point to ensure uniform detector acceptance. Beam-gas events are rejected using offline timing requirements with the V0 detector. The interaction probability per single bunch crossing was below 0.01 (5$\times$10$^{-3}$) during the entire data taking period at \s = 5.02 TeV (13 TeV). The residual contamination with pile-up events is rejected offline using an algorithm which identifies multiple primary vertices reconstructed with SPD tracklets and global tracks~\cite{Abelev:2014ffa}. 

Selected tracks are required to have a minimum transverse momentum of 1 \GeVc, a pseudorapidity in the range of $|\eta| < 0.9$, a minimum of 70 space points in the TPC, and a value of the track fit $\chi^{2}$ over the number of track points smaller than 4. 
A hit in at least one of the two SPD layers is also required to improve the tracking resolution, reduce the number of electrons from photon conversion in the detector material, and suppress tracks from pile-up collisions occurring in different bunch crossings.  
In order to reject secondary tracks originating from weak decays and interactions with the detector material, the candidate tracks are also required to have a maximum distance-of-closest-approach (DCA) to the reconstructed collision vertex of  0.5 cm in the radial direction and 2.0 cm along the beam-axis direction. Tracks originating from topologically identified long-lived weak decays of charged pions or kaons 
are rejected from the analysis. The electron identification is done by requiring the reconstructed TPC \dEdx signal to lie within the interval [$-$2,+3] $\sigma_{\rm e}$ relative to the expectation for electrons, where $\sigma_{\rm e}$ is the specific energy-loss resolution for electrons in the TPC. Furthermore, tracks consistent with the pion and proton assumptions within 3.0 (3.5) $\sigma$ are rejected in pp collisions at \s = 5.02 TeV (13 TeV). 
In addition, at \s = 13 TeV the pion rejection was released from 3.5 to 2.5 $\sigma$ for tracks with a momentum larger than 6 \GeVc in order to increase the \pJPsi reconstruction efficiency in the highest \pt\ interval (10--15 GeV/$c$).
Finally, electrons, which are found to be compatible with electrons from gamma conversions when combined with an opposite charge candidate
selected with looser requirements, are rejected.

The \jpsi candidates are formed by considering all opposite charge electron pairs. Pair candidates where neither of the decay products has a hit in the first layer of the SPD are excluded for a \pt of the pair below 7 \GeVc due to the poor spatial resolution of the associated decay vertex. 
For higher values of the \pt of the pair, this condition is released to increase the number of candidates.
Prompt \pJPsi mesons are separated from those originating from beauty-hadron decays on a statistical basis, exploiting the displacement between the primary event vertex and the decay vertex of the \pJPsi. The measurement of the fraction of \pJPsi  mesons originating from beauty-hadron decays, \fb, is carried out through an unbinned two-dimensional likelihood fit procedure, following the same technique adopted in the previous pp analysis~\cite{Abelev:2012gx}. A simultaneous fit of the dielectron pair invariant mass (\mee) and pseudoproper decay length (\x) distribution is performed. The latter is defined as $x = c \times \Vec{L} \times \Vec{\pt} \times m_{\rm \pJPsi}/|\Vec{\pt}|$, where $\Vec{L}$ is the vector pointing from the primary vertex to the \pJPsi decay vertex and $m_{\rm \pJPsi}$ is the \pJPsi mass provided by the Particle Data Group (PDG)~\cite{Tanabashi:2018oca}. The fit procedure maximises the logarithm of a likelihood function:

\begin{equation}
\label{eq:likeFunc} 
\ln{\mathcal{L}}= \sum_{i=1}^{N} \ln{\left[ \fsig \times F_{\rm Sig}(\x^{i}) \times M_{Sig}(\mee^{i}) + (1 - \fsig) \times F_{\rm Bkg}(\x^{i}) \times M_{\rm Bkg}(\mee^{i})\right]},
\end{equation}

where $N$ is the number of \pJPsi candidates within the invariant-mass interval $2.4 < \mee < 3.6$ GeV/$c^{2}$, $F_{\rm Sig}(\x)$ and $F_{\rm Bkg}(\x)$ represent the \pt-dependent probability density functions (PDFs) for the pseudoproper decay length distributions of signal and background, respectively. Similarly, $M_{Sig}(\mee)$ and $M_{\rm Bkg}(\mee)$ represent the equivalent PDFs for the invariant mass distributions. The signal fraction within the invariant-mass window considered for the fit, \fsig, represents the relative fraction of signal candidates, both prompt and non-prompt, over the sum of signal and background. The pseudoproper decay length PDF of the signal is defined as: 

\begin{equation}
\label{eq:signalPseudoPropDL} 
F_{\rm Sig}(x) =  \fbp \times F_{\rm B}(x) + (1-\fbp) \times F_{\rm prompt}(x),
\end{equation}

where $F_{\rm B}(x)$ and $F_{\rm prompt}(x)$ are the $x$ PDFs for non-prompt and prompt \pJPsi, respectively while \fbp\ represents the fraction of \pJPsi originating from beauty-hadron decays retrieved from the maximum likelihood fit procedure. The only free parameters in the fitting procedure are \fsig\ and \fbp. The latter needs to be corrected for the different acceptance-times-efficiencies for prompt and non-prompt \pJPsi, averaged in the \pt\ range where the measurement is performed. These differences can arise from the different \pJPsi \pt\ distributions used in the simulations, as well as different polarisation, as described below. The fraction of non-prompt \pJPsi corrected for these effects, \fb, is obtained as:

\begin{equation}
\label{eq:fBCorrProcedure} 
\fb = \left( 1 + \frac{1-\fbp}{\fbp} \times \frac{\langle A \times \epsilon \rangle_{\rm B~~~~~~}}{\langle A \times \epsilon \rangle_{\rm prompt}} \right)^{-1},
\end{equation}

where $\langle A \times \epsilon \rangle_{\rm prompt}$ and $\langle A \times \epsilon \rangle_{\rm B}$ represent the average acceptance-times-efficiency values for prompt and non-prompt \pJPsi, respectively, in the considered \pt\ interval.  

\begin{figure}[h!]
  \begin{center}
  \includegraphics[width=0.94\textwidth]{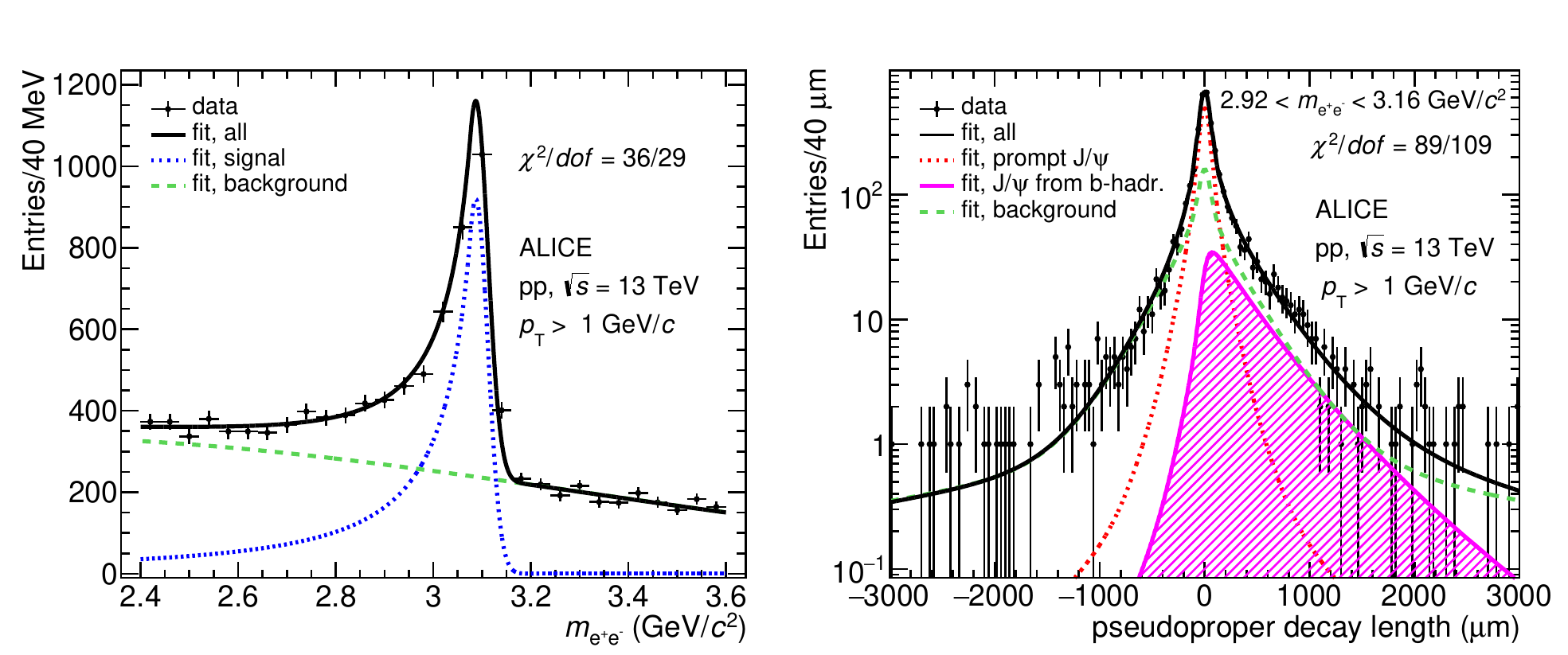} \\
  \includegraphics[width=0.94\textwidth]{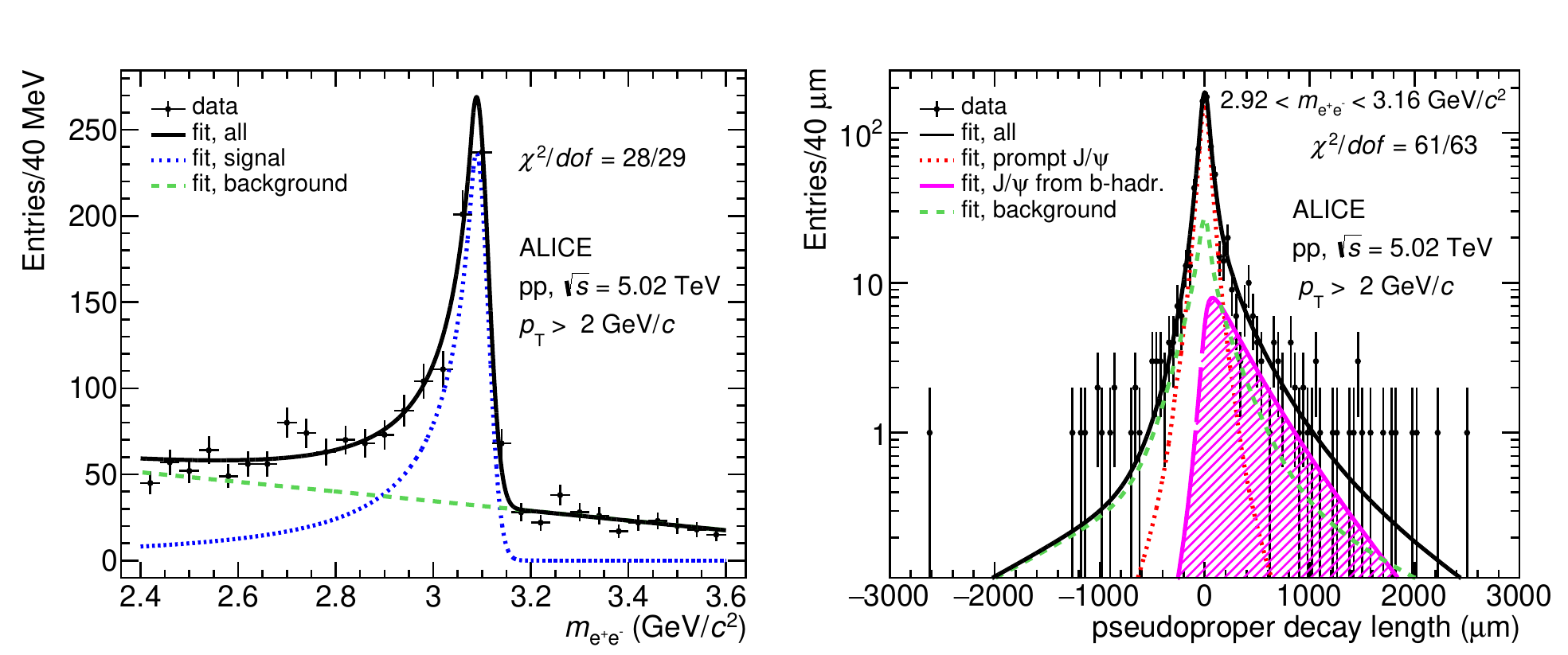}
  \end{center}
  \caption{Invariant mass (left panels) and pseudoproper decay length (right panels) distributions for \pJPsi candidates at midrapidity with superimposed projections of the maximum likelihood fit. Pseudoproper decay length distributions are shown for \pJPsi  candidates reconstructed under the \pJPsi mass peak, i.e. for $2.92 < \mee < 3.16$ \GeVmass, for display purposes only. The distributions refer to the \pt-integrated case, in particular the upper (lower) panels show the distributions for \pt\ $>$ 1 (2) \GeVc in pp collisions at \s = 13 (5.02) TeV. The $\chi^{2}$ values, which are computed considering the binned distributions of data points and the corresponding projections of the total fit function, are also reported. }
  \label{Fig:likelihoodFits}
\end{figure}

The various PDFs entering into the determination of \fb\ are described in Refs.~\cite{Abelev:2012gx,Adam:2015rba}. The PDFs corresponding to \pJPsi, namely $F_{\rm prompt}(x)$, $F_{\rm B}(x)$ and  $M_{Sig}(\mee)$, are determined from Monte Carlo (MC) simulations. A sample of minimum bias pp collisions is generated using PYTHIA 6.4~\cite{Sjostrand:2006za}. To this sample, prompt and non-prompt \pJPsi mesons are added. The latter is also generated with PYTHIA 6.4, while the prompt \pJPsi are simulated with a \pt\ spectrum based on a phenomenological interpolation
of measurements at RHIC, CDF, and the LHC~\cite{Bossu:2011qe} and a uniform distribution in rapidity. The \pJPsi dielectron decay is simulated with the EvtGen~\cite{Lange:2001uf} package, using the PHOTOS model~\cite{BARBERIO1994291} to deal with the influence of radiative decays ($\pJPsi \rightarrow e^{+}e^{-}\gamma$). Finally, GEANT3~\cite{Brun:1082634} is employed to handle the particle transport through the ALICE apparatus, considering a detailed description of the detector material and geometry. Detector responses and calibrations in MC simulations are tuned to data, taking into account time dependent running conditions of all detectors included in the data acquisition.
These simulations are used to compute the corrections for acceptance and efficiency for the corresponding inclusive \pJPsi cross section analyses at \s = 5.02 and 13 TeV,
as described in Refs.~\cite{Acharya:2019lkw,Alice13TeV}. 
In the aforementioned MC simulations prompt \pJPsi component is assumed unpolarised, whereas for the non-prompt \pJPsi a small
residual polarisation as predicted by EvtGen~\cite{Lange:2001uf} is considered. Further assumptions concerning the polarisation of both components are not taken into account in this publication. In order to consider a realistic shape of the \pt\ distributions for evaluating the acceptance-times-efficiency corrections averaged over finite size \pt intervals, which enter in Eq.~\ref{eq:fBCorrProcedure}, the \pJPsi kinematic distributions used in simulations were tuned to match experimentally observed distributions. In particular, the measured inclusive cross section is used to reweight the MC \pt\ shape of prompt \pJPsi, whereas the reweighting of the non-prompt \pJPsi component is performed according to FONLL calculations. The largest difference between the average efficiencies of prompt and non-prompt \pJPsi is observed for the \pt-integrated case where it amounts to about 3\% at both $\sqrt{s}$ = 5 and 13 TeV.
One of the key ingredients is the  resolution function, R($x$), which describes the accuracy of \x\ in the reconstruction. It affects all PDFs in Eq.~(\ref{eq:likeFunc}) related to the pseudoproper decay length, and it is determined via MC simulations, considering the \x\ distributions of prompt \pJPsi\ reconstructed with the same procedure and selection criteria as for data.  It is described by the sum of two Gaussians and a symmetric power law function, and it was determined as a function of the \pt\ of the \pJPsi. Tuning of the MC simulations was applied to minimise the residual discrepancy between data and simulation for the distribution of the DCA in the transverse plane of single charged tracks, as done in Ref.~\cite{Acharya:2018yud}. The RMS of the resolution function in pp collisions at \s = 13 TeV for candidate pairs with both decay tracks having a hit in the first layer of the SPD, ranges from about 180~$\rm \mu m$ at \pt = 1.5 \GeVc to 40~$\rm \mu m$ at \pt = 12.5 \GeVc. The corresponding RMS of the resolution function for the 5.02 TeV data set is found to be about 30\% worse at similar \pt\ values compared to the 13 TeV case, mainly due to worse ITS detector performance. 
Background PDFs, both for invariant mass and pseudoproper decay length, are retrieved from data. In particular, the invariant-mass background PDF, $M_{\rm Bkg}(\mee)$, was parametrised by a second-order polynomial function for \pt\ below 2 GeV/$c$ and for the \pt-integrated case at \s = 13 TeV. For  \pt~$>$~2~GeV/$c$, an exponential function was used at both centre-of-mass energies. In particular, the parameters of the invariant mass background function were determined by fitting the invariant-mass distribution of opposite charge-sign pairs by $M_{\rm Bkg}(\mee)$ plus a Crystal Ball function~\cite{Gaiser:1982yw} for the signal, whose shape was determined from MC simulations. The background $x$ PDF considered in the two dimensional log-likelihood function defined by Eq.~\ref{eq:likeFunc}, ${\rm F}_{\rm Bkg}(x)$, was constrained by fitting the pseudoproper decay length distribution in the sidebands of the dielectron invariant-mass distribution, defined as the regions 2.4--2.6 and 3.2--3.6~GeV/$c^{2}$. 

Examples of ${\rm e}^{+}{\rm e}^{-}$ invariant mass and pseudoproper decay length distributions with superimposed projections of the total maximum likelihood fit functions, computed in the integrated \pt range, are shown in  Fig.~\ref{Fig:likelihoodFits} for 13 TeV (upper panels) and 5.02 TeV (lower panels) for \pt\ $>$ 1 and \pt\ $>$ 2 GeV/$c$, respectively.
Different components of the likelihood fit function are superimposed on the invariant-mass (left panels) and pseudoproper decay length (right panels) distributions of opposite charge-sign candidates.
In particular, for the pseudoproper decay length the components relative to the background, prompt and non-prompt \pJPsi are shown. 
In addition to the \pt-integrated case, the fraction of non-prompt \pJPsi was also studied in six intervals of transverse momentum (1--2, 2--3, 3--5, 5--7, 7--10, 10--15 \GeVc) at \s = 13 TeV and three \pt\ intervals (2--4, 4--7, 7--10 \GeVc) at \s = 5.02 TeV. Furthermore, the data sample collected at \s = 13 TeV enables the study of the non-prompt \pJPsi fraction differentially in rapidity; in particular it was measured in three rapidity intervals for \pt\ $>$ 1 \GeVc (\mbox{$|y| < 0.2$,} \mbox{$0.2 < |y| < 0.5$,} \mbox{$0.5 < |y| < 0.9$}). 

Systematic uncertatinties on the non-prompt fraction originate from the uncertainty on the $\langle A \times \epsilon \rangle$, through Eq.~\ref{eq:fBCorrProcedure}, as well as incomplete knowledge of the PDFs. Systematic uncertainties related to different hypotheses of polarisation of prompt and non-prompt \pJPsi are not considered.

The systematic uncertainty on $\langle A \times \epsilon \rangle$ arising from the variation of the selection criteria is described in detail in Ref.~\cite{Alice13TeV}. 
The dominant sources of uncertainty are related to the tracking and electron identification procedures. However, these are expected to be fully correlated for prompt and non-prompt \pJPsi, 
therefore their contribution cancels out in the ratio when propagated to \fb\ according to Eq.~\ref{eq:fBCorrProcedure}.  
The systematic uncertainty of $\langle A \times \epsilon \rangle$ obtained on the non-prompt \pJPsi fraction is evaluated by repeating
the estimation of the ratio $\frac{\langle A \times \epsilon \rangle_{\rm B~~~~~~}}{\langle A \times \epsilon \rangle_{\rm prompt}}$,
considering different hypotheses for transverse momentum distributions of prompt and non-prompt \pJPsi.
For non-prompt \pJPsi, reweighted according to FONLL, different \pt\ distributions obtained from varying FONLL
parameters are considered. For prompt \pJPsi, the measured \pJPsi cross section is fitted and the envelope obtained by varying the fitting parameters according to their
uncertainties is considered. The systematic uncertainty on \fb\ is assigned by considering the maximum difference observed when varying the aforementioned ratio.

Systematic uncertainties, originating from the incomplete knowledge of all PDFs employed in the fitting procedure, are determined following a similar approach as described in previous ALICE analyses~\cite{Abelev:2012gx,Adam:2015rba,Acharya:2018yud}. An additional contribution considered in the analyses presented in this article is related to the uncertainty of the relative hadronisation fractions of beauty quarks into beauty hadrons. The mixture of beauty hadrons in MC simulations can affect the PDF used for the description of the pseudoproper decay length distribution of non-prompt \pJPsi. Beauty-hadron fractions in PYTHIA 6.4 are simulated uniformly in \pt, with the corresponding values compatible with those from the PDG~\cite{Tanabashi:2018oca}. The LHCb collaboration measured the production fractions of $\overline{\rm B}^{\rm 0}_{\rm s}$ and ${\rm \Lambda}^{\rm 0}_{\rm b}$ hadrons, normalised to the sum of ${\rm B}^{-}$ and $\overline{\rm B}^{0}$ mesons, at forward rapidity ($2 < \eta < 5$) in pp collisions at \s = 13 TeV~\cite{PhysRevD.100.031102}. It was found that the ${\rm \Lambda}^{\rm 0}_{\rm b}$ to (${\rm B}^{-}+\overline{\rm B}^{0}$) ratio depends strongly on the transverse momentum of the beauty-hadron, in particular it is about 0.12 at \pt\ = 25 \GeVc and it increases significantly at low transverse momentum, reaching about 0.3 at \pt\ = 4 \GeVc, 
and showing no dependence on pseudorapidity. The relative fractions of beauty hadrons in MC simulations, employed to obtain the pseudoproper decay length PDF of non-prompt \pJPsi, were thus reweighted in order to match those measured by the LHCb collaboration. The corresponding systematic uncertainty was assigned by considering half of the relative deviation obtained on \fb\ when MC simulations without the reweighting procedure are used.

The systematic uncertainties were studied for each individual \pt interval, as well as for the \pt-integrated case. In pp collisions at \s = 13 TeV no significant rapidity dependence of the systematic uncertainties was observed, therefore the systematic uncertainties assigned in the three rapidity intervals are the same as those evaluated for the $y$-integrated case.

\begin{table} [t!]
\centering
\caption{Systematic uncertainties of \fb, expressed in \%, for all \pt\ intervals considered in the analysis performed at \s = 13 TeV. In the last row the statistical uncertainties on \fb\ are also reported.}
\begin{tabular}{ |c|c|c|c|c|c|c|c| }
 \hline
 
                 &  \multicolumn{7}{c|}{\pt\ (GeV/$c$)}  \\
                
                 &  $>$ 1 & 1--2 & 2--3 & 3--5 & 5--7 & 7--10 & 10--15  \\
 \hline
	Resolution function R($x$)&   4.0 & 10.9 & 5.6 & 3.1 & 1.5 & 0.9 & 0.7 \\
\x\ PDF of background &  4.1 & 9.4 & 7.6 & 3.1 & 2.8 & 2.0 & 2.9 \\
 \x\ PDF of non-prompt \pJPsi & 3.3 & 5.5 & 4.3 & 2.5 & 1.5 & 0.8 & 0.6 \\
 Primary vertex & 2.5 & 4.4 & 4.3 & 2.3 & 1.2 & 0.5 & 0.3 \\ 
MC \pt\ distribution &2.6 & 0.5 & 0.5 & 0.5 & 0.1 & 0.1 & 0.1 \\
	$m_{\rm ee}$ PDF of signal & 0.5 & 0.7 & 0.4 & 0.4 & 0.4 & 0.4 & 0.4 \\
 $m_{\rm ee}$ PDF of background & 2.1 & 0.5 & 0.8 & 1.1 & 1.1 & 1.1 & 1.1 \\
 \hline
 Total & 7.8 & 16.0 & 11.2 & 5.7 & 3.9 & 2.6 & 3.2 \\
 \hline
 \hline
 Stat. uncertainty & 8.1 & 33.3 & 18.7 & 12.7 & 12.7 & 12.9 & 20.2 \\
 \hline
 \end{tabular}
\label{tab:fbsystable13TeV}
\end{table}

\begin{table} [t!]
\centering
\caption{Systematic uncertainties of \fb, expressed in \%, for all \pt\ intervals considered in the analysis performed at \s = 5.02 TeV. In the last row the statistical uncertainties on \fb\ are also reported.}
\begin{tabular}{ |c|c|c|c|c|c| }
 \hline
 
          &  \multicolumn{4}{c|}{\pt\ (GeV/$c$)}  \\
         &  $>$ 2 & 2--4 & 4--7 & 7--10  \\
 \hline
Resolution function R($x$) & 3.0 & 4.4 & 1.6 & 0.6 \\
\x\ PDF of background & 6.4 & 11.2 & 3.4 & 3.5\\
 \x\ PDF of non-prompt \pJPsi &  3.1 & 4.2 & 1.8 & 1.5 \\
  Primary vertex & 5.0 & 7.8 & 3.3 & 1.5 \\
  MC \pt\ distribution & 3.0 & 1.0 & 0.5 & 0.5  \\
  $m_{\rm ee}$ PDF of signal & 0.6 & 0.7 & 0.7 & 0.7 \\
 $m_{\rm ee}$ PDF of background &  1.0 & 1.5 & 1.0 & 1.0 \\
 \hline
 Total &   9.7 & 15.0 & 5.4 & 4.4  \\
 \hline
 \hline
 Stat. uncertainty &   14.6 & 23.2 & 19.5 & 29.2 \\
 \hline
 \end{tabular}
\label{tab:fbsystable5TeV}
\end{table}

Systematic uncertainties are summarised for the \pt-integrated case, as well as in transverse momentum intervals, in Tables~\ref{tab:fbsystable13TeV} and~\ref{tab:fbsystable5TeV} for pp collisions at \s = 13 and 5.02 TeV, respectively. The largest contributions to the total systematic uncertainty come from the resolution function and the PDF of the pseudoproper decay length for the background. 
A large contribution is associated to the reconstruction of the primary vertex, which might include in its computation the
decay tracks of the \pJPsi candidates, both prompt and non-prompt. 
A systematic uncertainty to account for possible bias effects due to the presence of decay products from non-prompt \pJPsi candidates was evaluated, similarly as done in the previous analyses at \s = 7 TeV~\cite{Abelev:2012gx,Adam:2015ota}. The systematic uncertainty related to the primary vertex was found to be larger in the 5.02 TeV data set than in the 13 TeV one due to the lower multiplicity.
The systematic uncertainty on the reconstruction of the primary vertex in Table~\ref{tab:fbsystable13TeV} and Table~\ref{tab:fbsystable5TeV} shows an increasing trend towards lower transverse momentum at both centre-of-mass energies, reaching about 8\% in the \pt\ interval 2--4 \GeVc at \s = 5.02 TeV. Both the larger systematic uncertainty of the primary vertex and the  worse \x\ resolution at \s = 5.02 TeV than at \s = 13 TeV, lead to the choice of  2 \GeVc as the lower \pt\ threshold for reconstructed \pJPsi candidates at \s = 5.02 TeV, whereas the non-prompt \pJPsi fraction could be determined down to 1 \GeVc at \s = 13 TeV. The total systematic uncertainty was obtained at both centre-of-mass energies by adding in quadrature the contributions from all sources detailed in Tables~\ref{tab:fbsystable13TeV} and~\ref{tab:fbsystable5TeV}. Most of the systematic sources can be considered highly correlated over the \pt\ ranges, except those related to the $x$ and $m_{\rm ee}$ PDF of the background.

\section{Results and discussion}
\label{Sec:results}

\begin{figure}[t!]
  \includegraphics[width=0.49\textwidth]{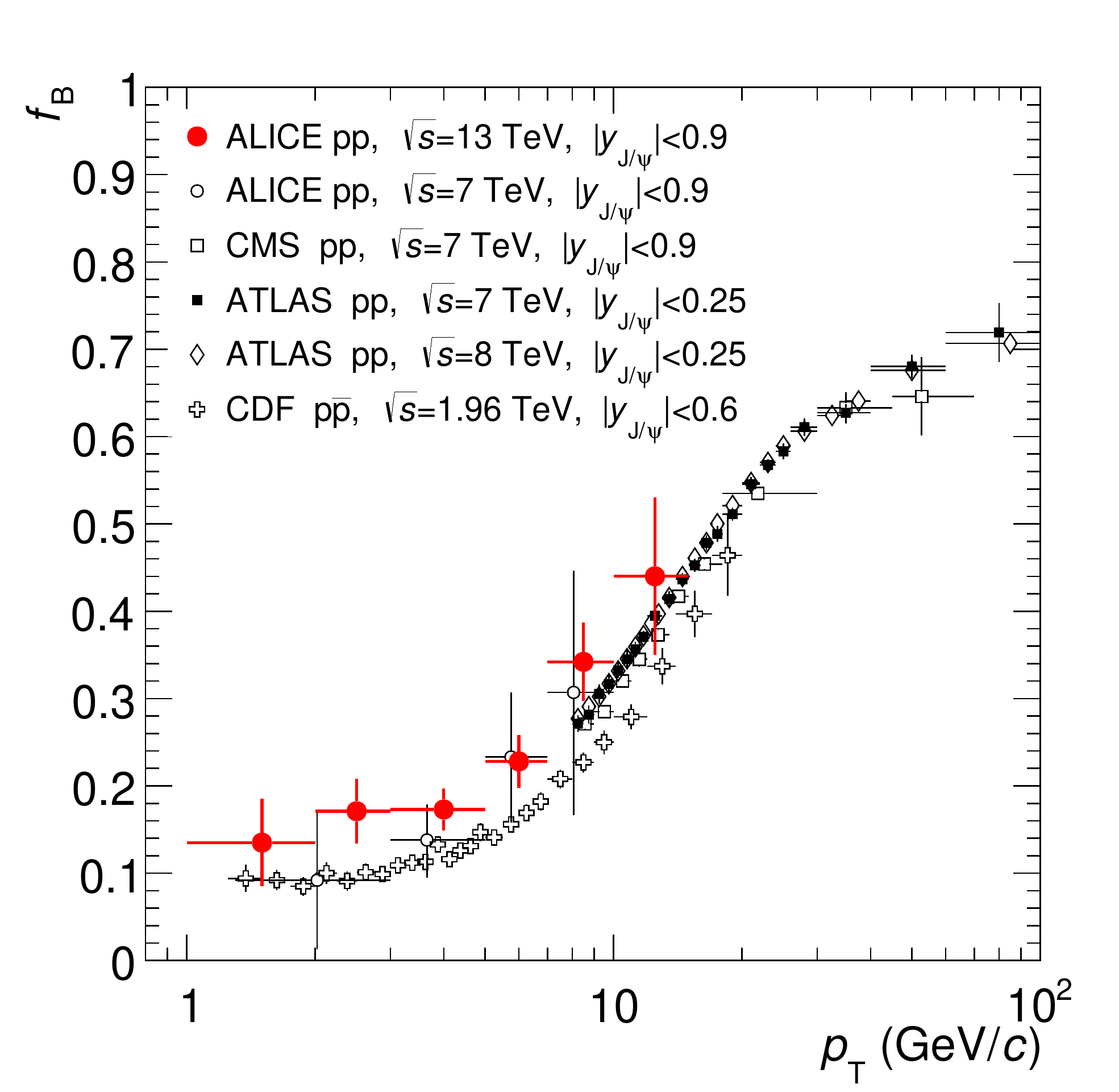}
  \includegraphics[width=0.49\textwidth]{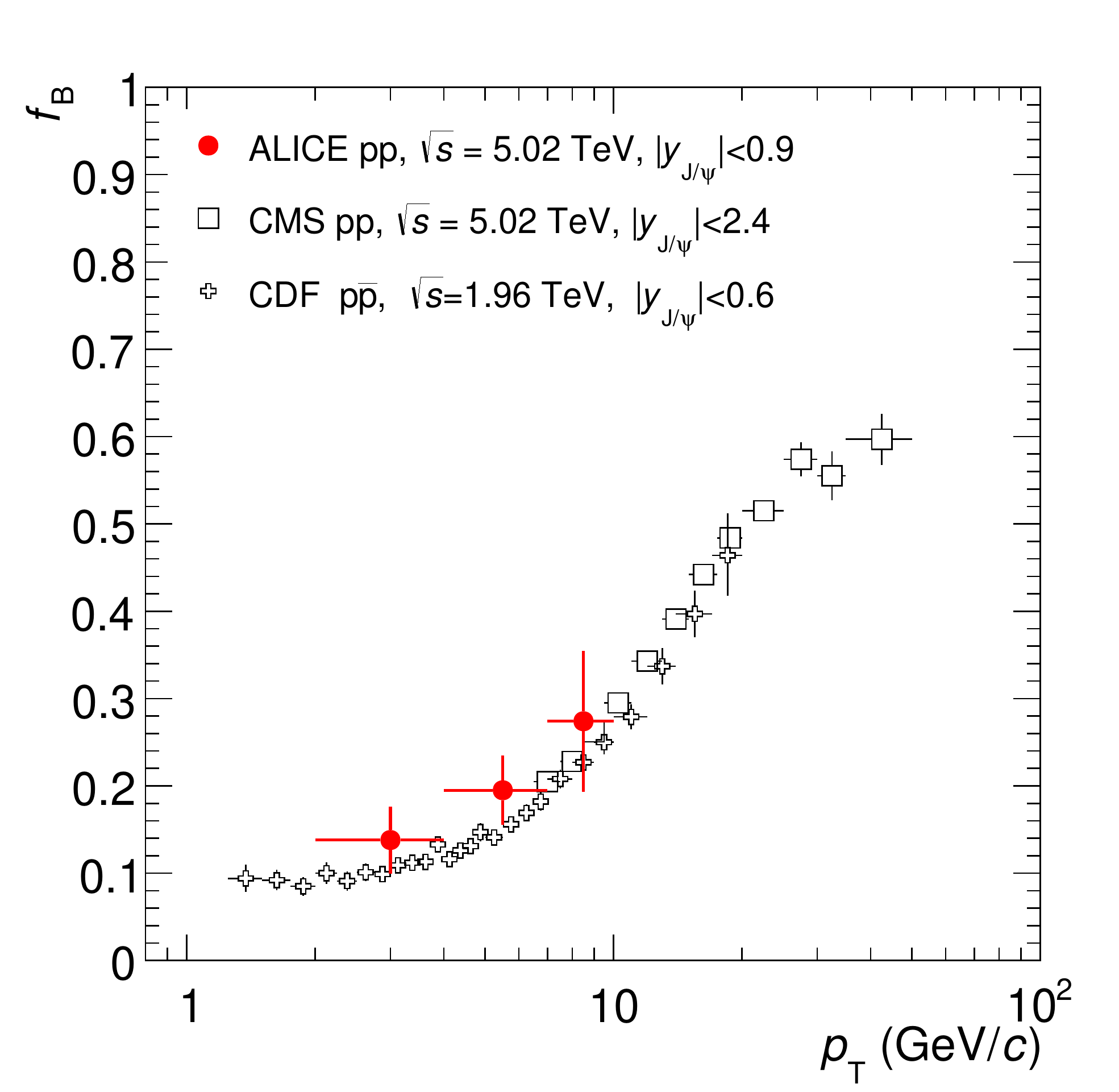}
	\caption{Non-prompt \pJPsi fraction as a function of \pt measured by the ALICE collaboration in pp collisions at \s = 13 TeV (left panel) and 5.02 TeV (right panel) compared with similar results obtained at midrapidity in \pp collisions at the LHC, namely CMS~\cite{Khachatryan:2010yr,Sirunyan:2017isk} and ATLAS~\cite{ATLAS:2015zdw}. The results from CDF in proton\textendash antiproton collisions at \s = 1.96 TeV~\cite{PhysRevD.71.032001} are also shown in both panels. Error bars correspond to the quadratic sum of statistical and systematic uncertainties.
	}
  \label{Fig:fbResults}
\end{figure}

Figure~\ref{Fig:fbResults} shows the non-prompt \pJPsi fraction measured by the ALICE Collaboration in pp collisions at \s = 13 TeV (left panel) and 5.02 TeV (right panel) as a function of transverse momentum, compared to the measurements carried out at the LHC at midrapidity by the CMS~\cite{Khachatryan:2010yr,Sirunyan:2017isk} and the ATLAS~\cite{ATLAS:2015zdw} collaborations. The measurements performed by the CDF Collaboration in proton\textendash antiproton collisions at \s = 1.96 TeV~\cite{PhysRevD.71.032001} are also shown in both panels. 
The non-prompt \pJPsi fractions measured by the ALICE collaboration exhibit an increasing trend as a function of the transverse momentum of the \pJPsi mesons, inline with previously published measurements. 
The ALICE results at \s~=~5.02~TeV are compatible with those from CMS~\cite{Sirunyan:2017isk} in the common \pt range. 
A comparison of the ALICE results at \s = 13 TeV and the lower centre-of-mass energy measurements from CDF hints that the 
increase of \fb\ with collision energy, previously observed from the ATLAS and CMS measurements, holds also at lower \pt.

The fractions of \pJPsi originating from beauty-hadron ($h_\mathrm{B}$) decays within $|y| < 0.9$ in pp collisions at \s~=~13~and~5.02~TeV in the measured \pt\ intervals, 
also called the ``visible'' regions, are: 

\begin{align*}
	\fb^{{\rm visible,}~\s~=~{\rm 13~TeV}}(\pt > {\rm 1~GeV}/{\it c}, |y| < 0.9) &= 0.185 \pm 0.015~(\rm stat.) \pm  0.014~(\rm syst.), \\
	\fb^{{\rm visible,}~\s~=~{\rm 5.02~TeV}}(\pt > {\rm 2~GeV}/{\it c}, |y| < 0.9) &= 0.157 \pm 0.023~(\rm stat.) \pm 0.015~(\rm syst.).
\label{eq:fbresults} 
\end{align*}

The non-prompt \pJPsi\ fractions can be combined with the corresponding inclusive \pJPsi\ cross sections measured at the two centre-of-mass energies~\cite{Acharya:2019lkw,Alice13TeV} to evaluate prompt and non-prompt \pJPsi cross sections. The values provided in this section are quoted assuming prompt \pJPsi to be unpolarised, whereas for non-prompt \pJPsi
a residual polarisation, as described in Section~\ref{Sec:dataAnalysis}, is considered. Non-prompt and prompt \pJPsi cross sections are obtained according to:

\begin{equation}
   \sigma_{\pJPsi \leftarrow {\it h}_{\rm B}} = \fb \times \sigma_{\rm inclusive~\pJPsi},~~~~~~~~~~~~~~~~\sigma_{\rm prompt~\pJPsi} = (1-\fb) \times \sigma_{\rm inclusive~\pJPsi}. 
        \label{eq:fbAndInclCombination}
\end{equation}

\begin{figure}[t!]

  \includegraphics[width=0.49\textwidth]{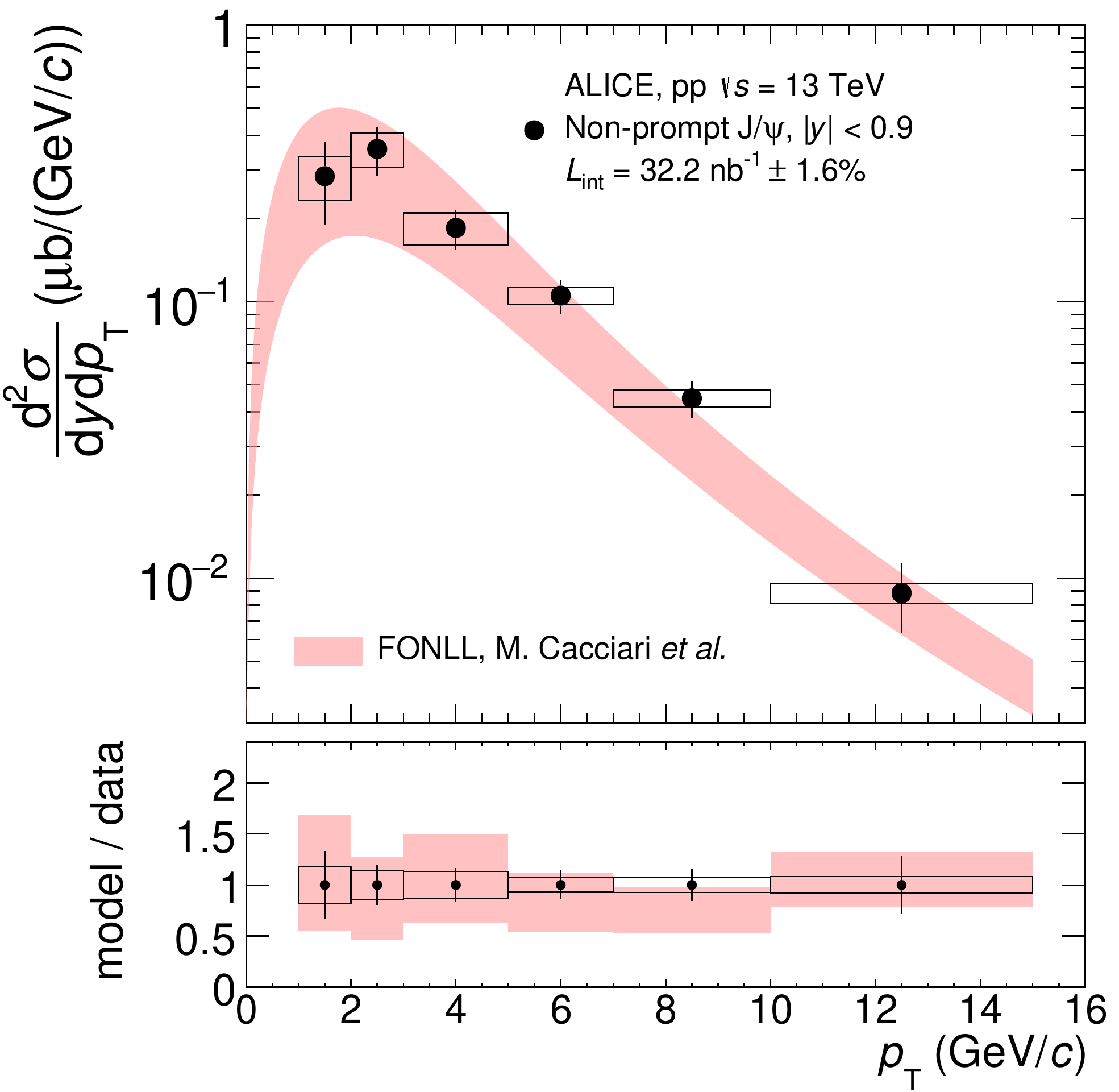}
  \includegraphics[width=0.49\textwidth]{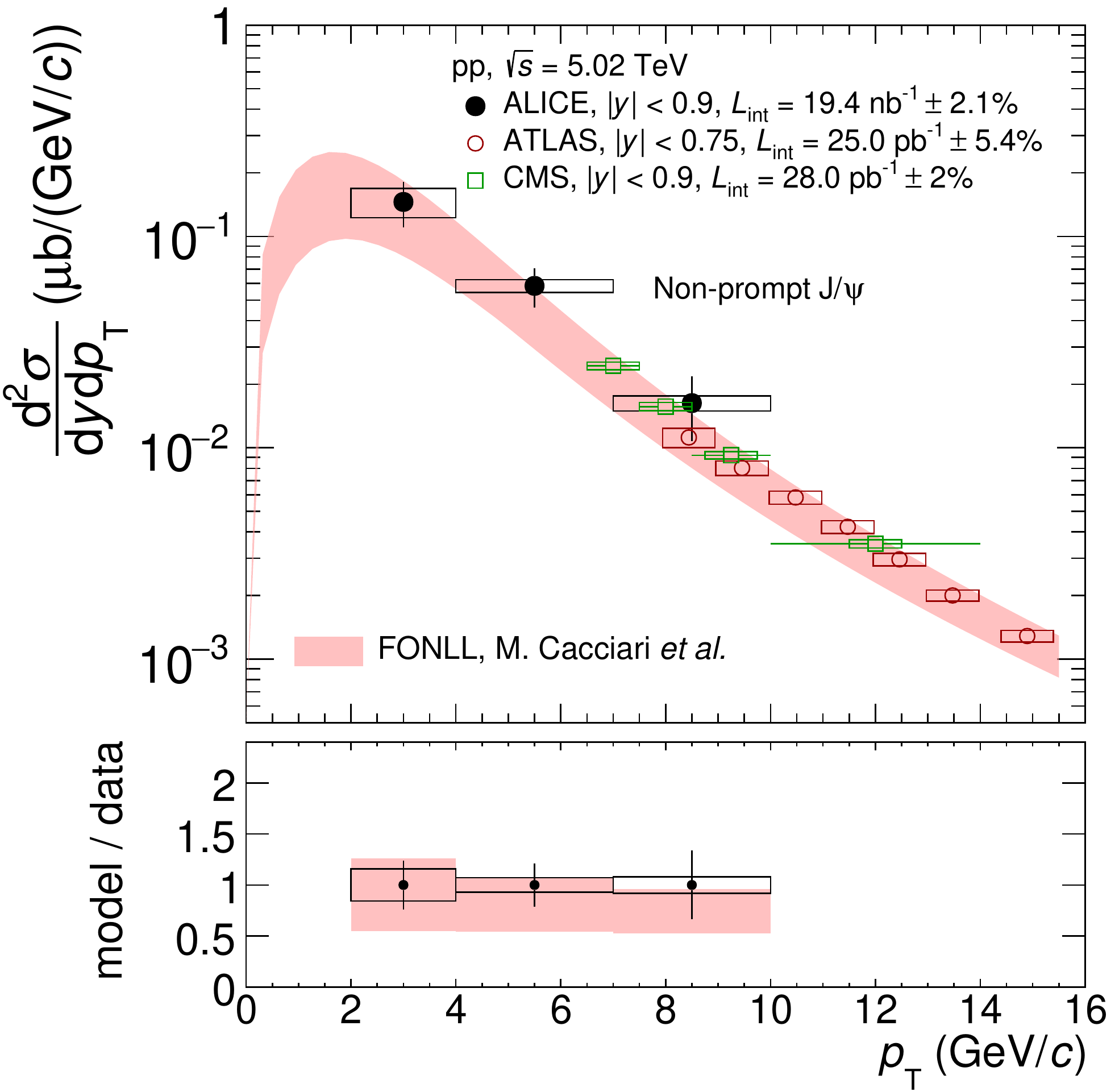}
	\caption{The $\frac{{\rm d}^2\sigma}{{\rm d}y {\rm d}\pt}$ of non-prompt \pJPsi\ in pp collisions at $\sqrt{s}$ = 13 TeV (left panel) and 5.02 TeV (right panel) as a function of \pt. 
	 The measurement at \s = 5.02 TeV is compared with similar measurements from CMS~\cite{Sirunyan:2017mzd} and ATLAS~\cite{Aaboud:2017cif} collaborations at high \pt.
  The error bars (boxes) represent the statistical (systematic) uncertainties. Uncertainties due to the luminosity 
	are not included in the boxes. The results are compared with the FONLL calculations~\cite{Cacciari_2001,Cacciari:2012ny} at both energies.  
        Bottom panels show the ratios FONLL to ALICE data. The uncertainty band represents the relative uncertainty from the model whereas the bars and boxes centered around 
	unity refer to relative statistical and systematic uncertainties on ALICE data points, respectively. }
  \label{Fig:NonPromptXsecResults}
\end{figure}

The \pt-differential non-prompt \pJPsi cross sections at \s = 13 and 5.02 TeV are shown in the upper panels of Fig.~\ref{Fig:NonPromptXsecResults}.
Statistical and systematic uncertainties on the non-prompt \pJPsi cross section, shown in Fig.~\ref{Fig:NonPromptXsecResults} by error bars and boxes respectively, are evaluated
by adding in quadrature the corresponding uncertainties of \fb\ and inclusive \pJPsi\ cross sections.
Boxes do not include the global normalisation uncertainty due to the luminosity. The measurement at \s = 5.02 TeV is compared with the existing midrapidity results at higher \pt\ from ATLAS~\cite{Aaboud:2017cif} and CMS~\cite{Sirunyan:2017mzd}, at the same centre-of-mass energy. Consistency is observed with both the ATLAS and CMS measurements in the common \pt\ region.
These measurements are compared with theoretical calculations based on the FONLL factorisation approach~\cite{Cacciari_2001,Cacciari:2012ny}. For this calculation CTEQ6.6~\cite{Nadolsky:2008zw} parton distribution functions are used. The theoretical uncertainties from the factorisation and renormalisation scales, $\mu_{\rm F}$ and $\mu_{\rm R}$, are estimated by varying them independently in the ranges 0.5 $< \mu_{\rm F}/m_{\rm T} <$ 2 and 0.5 $< \mu_{\rm R}/m_{\rm T} <$ 2, with the constraint 0.5 $< \mu_{\rm F}/\mu_{\rm R} <$ 2 and $m_{\rm T}$ = $\sqrt{p_{\rm T}^{2} + m_{\rm b}^{2}}$. The beauty-quark mass was varied within 4.5 $< m_{\rm b} <$ 5.0 GeV/$c^2$. The uncertainties of the parton distribution functions, calculated according to the Hessian prescription of CTEQ6.6~\cite{Nadolsky:2008zw}, are included as well in the total uncertainty. The ratios of the model predictions to the ALICE data
are shown in the bottom panels at both energies. The relative uncertainties of the FONLL calculations are shown by the shaded band, while the data points around unity show the 
relative statistical and systematic uncertainties of the cross sections measured by the ALICE collaboration. 
 Most of the ALICE data points at \s = 5.02 and 13 TeV sit in the middle or upper regions of the corresponding FONLL uncertainty band, thus experimental results and theoretical calculations are compatible, albeit the theoretical uncertainties are significantly larger than the experimental ones, especially at low \pt.

\begin{figure}[t!]
  \includegraphics[width=0.49\textwidth]{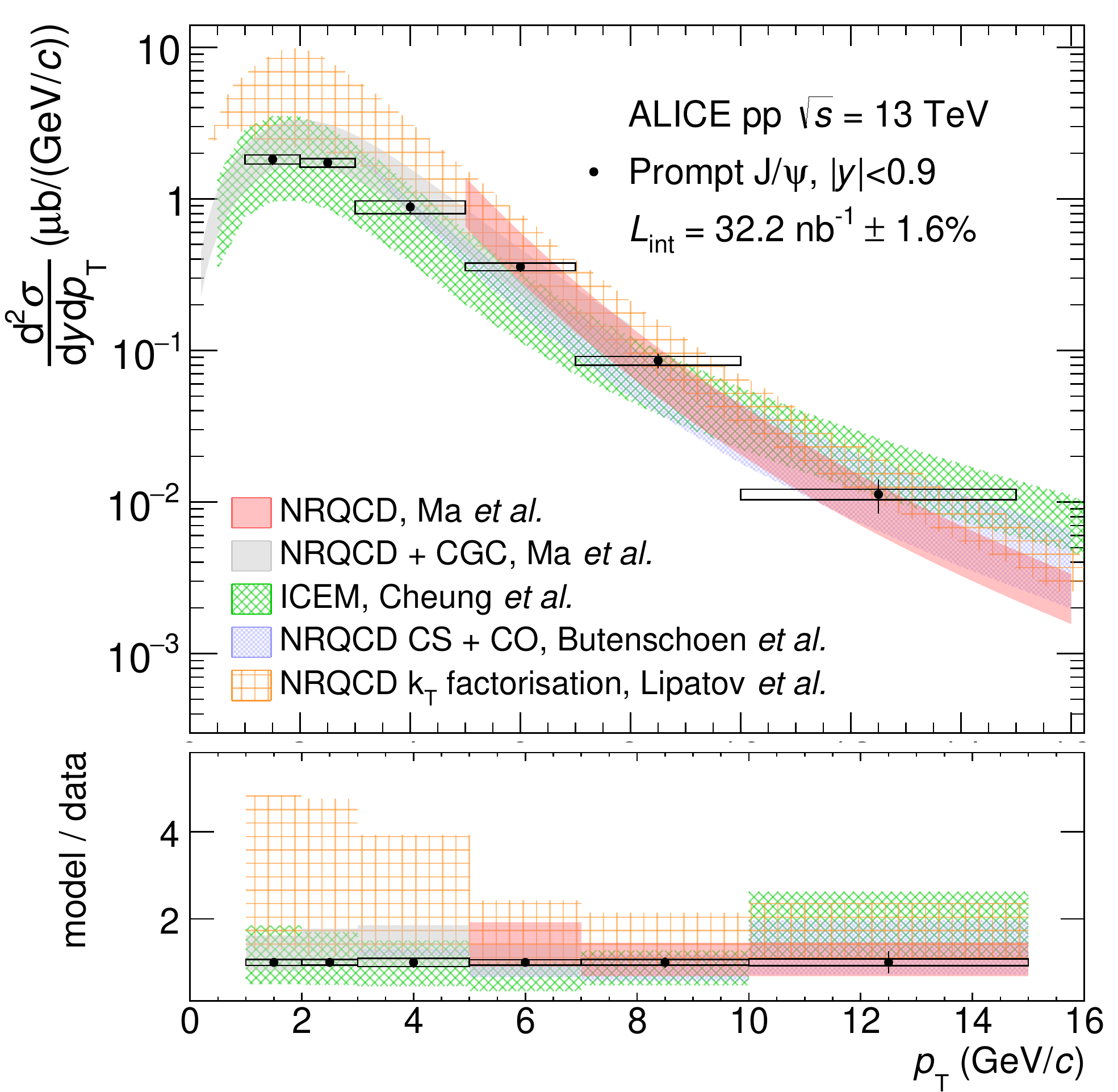} 
  \includegraphics[width=0.49\textwidth]{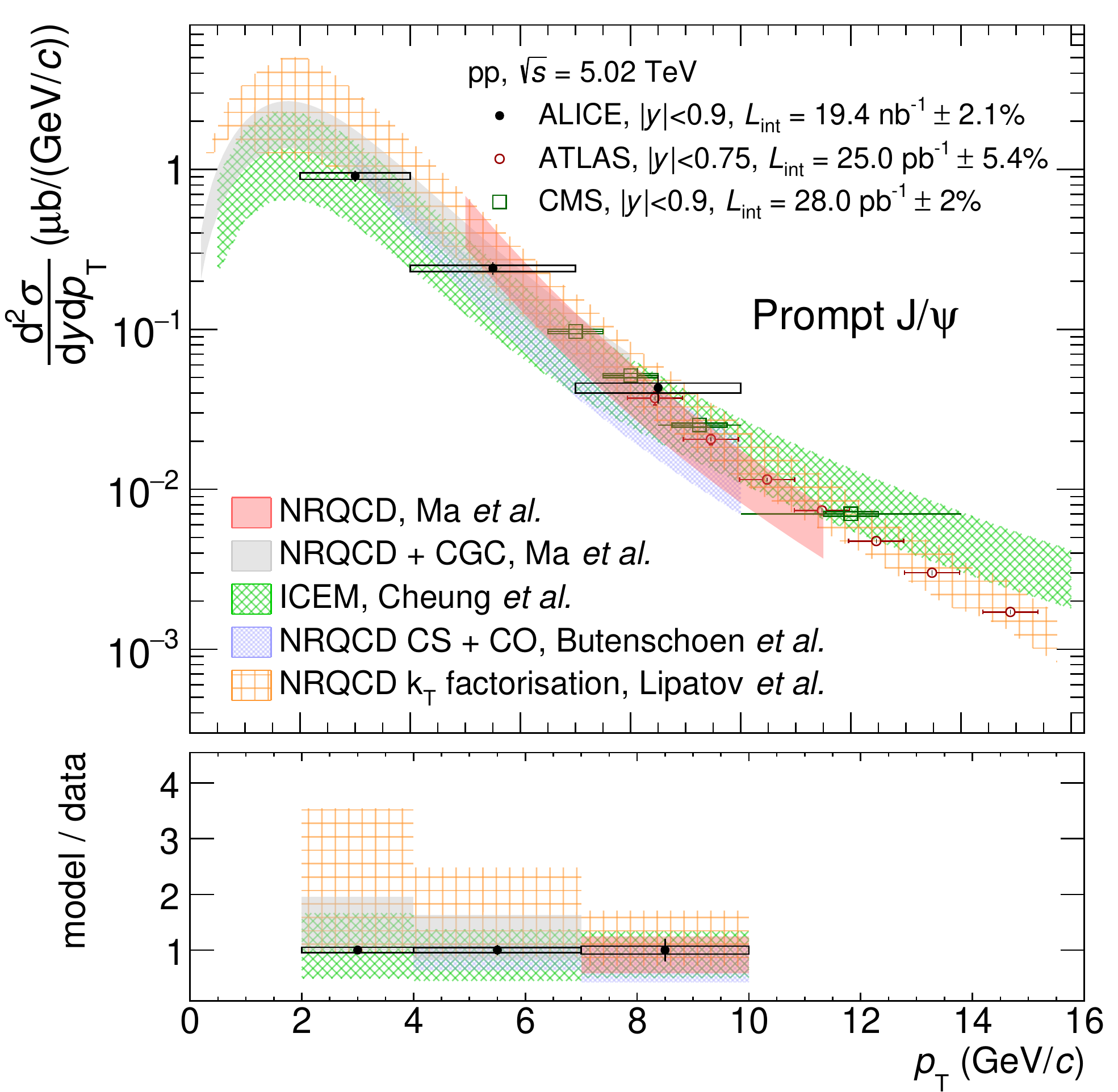}
  \caption{The $\frac{{\rm d}^2\sigma}{{\rm d}y {\rm d}\pt}$ of prompt \pJPsi measured by the ALICE collaboration in pp collisions at \s = 13 TeV (left panel) and \s = 5.02 TeV (right panel).  The error bars (boxes) represent the statistical (systematic) uncertainty.  
  The results at \s = 5.02 TeV are compared with similar measurements from CMS~\cite{Sirunyan:2017mzd} and ATLAS~\cite{Aaboud:2017cif} at high \pt.  
  Uncertainties due to the luminosity 
  are not included in the boxes. The results are compared with calculations from NLO NRQCD~\cite{Ma:2010yw,Butenschoen:2010rq,Baranov:2019lhm}, NRQCD+CGC~\cite{Ma:2014mri}, and from ICEM~\cite{PhysRevD.98.114029}. 
    Bottom panels show the ratios of the models to ALICE results. The uncertainty bands represent the relative uncertainty from each model whereas the bars and boxes centered around
        unity refer to relative statistical and systematic uncertainties on ALICE data points, respectively.} 
 \label{Fig:PromptXsecResults}
\end{figure}

The \pt-differential cross section of prompt \pJPsi, also obtained according to Eq.~\ref{eq:fbAndInclCombination}, is shown in Fig.~\ref{Fig:PromptXsecResults} for both collision energies. A comparison of the 5.02 TeV measurement with the available measurements from ATLAS~\cite{Aaboud:2017cif} and CMS~\cite{Sirunyan:2017mzd} at midrapidity at the same energy shows consistency in the common \pt range. These measurements are also compared with theoretical calculations performed for both energies using a few NRQCD based models and the improved CEM model. In particular, the ratios of different models to ALICE measurements, with the corresponding relative uncertainties from each model, are shown in the bottom panels of Fig.~\ref{Fig:PromptXsecResults}. The NRQCD calculations by Ma {\it et al.}~\cite{Ma:2010yw} and Butenschoen {\it et al.}~\cite{Butenschoen:2010rq} are performed at NLO using collinear factorisation while the calculations by Ma and Venugopalan~\cite{Ma:2014mri} are leading order NRQCD calculations combined with a resummation of soft gluons within the Color-Glass Condensate (CGC) model. 
The two NLO calculations use different long distance matrix elements (LDME), obtained by fitting different charmonium measurements and in different kinematic intervals which leads to different \pt intervals of applicability. In addition, the calculations from Ref.~\cite{Butenschoen:2010rq} do not consider the contributions from decays of higher mass charmonia like $\psi$(2S) and $\chi_{\mathrm{c}}$, which are estimated to contribute more than 30\% to the prompt \pJPsi production~\cite{Faccioli:2008ir,LHCb:2012af}. Both calculations show good agreement with the data within the rather large theoretical uncertainties. 
The NRQCD+CGC calculations, which span over the whole transverse momentum region from \pt = 0 up to \pt = 8 \GeVc, show  good agreement at both centre-of-mass energies. 
The NLO NRQCD calculations by Lipatov {\it et al.}~\cite{Baranov:2019lhm}, obtained with the MC generator PEGASUS~\cite{Lipatov:2019oxs}, 
are performed within the $k_{\rm T}$-factorisation approach using \pt-dependent gluon distribution functions~\cite{Baranov:2019lhm}. 
The calculations can be extended down to zero transverse momentum of the J/$\psi$ using the KMR~\cite{Kimber:2001sc} technique to construct
the unintegrated gluon distribution functions. Furthermore, the LDMEs are obtained from a simultaneous fit of charmonium measurements at the LHC~\cite{Baranov:2019lhm}, and feed-down contributions to \pJPsi from higher charmonium states are taken into account.
The calculation overestimates the prompt \pJPsi production, especially in the low \pt region.  
The ICEM calculation from Cheung {\it et al.}~\cite{PhysRevD.98.114029}, performed within the $k_{\mathrm T}$-factorisation approach, provides a good description of the prompt \pJPsi cross section in the whole measured \pt range. This model includes feed-down contributions from higher mass charmonium states. 
The large model uncertainties exhibited by all the calculations are due to the unconstrained energy scales intrinsic to QCD calculations, namely the charm-quark mass, the renormalisation, and factorisation scales and affect mainly the overall normalisation of the calculations.

The integrated cross sections of \pJPsi\ from beauty-hadron decays in the visible regions at \s = 13 TeV and \s = 5.02 TeV are:

\begin{align*}
	\sigma_{\pJPsi \leftarrow {\it h}_{\rm B}}^{\rm visible,~ \sqrt{s} = 13 \rm ~TeV}(\pt > 1~{\rm GeV}/c, |y| < 0.9)~&=~2.71~\pm~0.23~(\rm stat.) \pm 0.25 (\rm syst.)~{\rm \mu b}, \\ 
\sigma_{\pJPsi \leftarrow {\it h}_{\rm B}}^{\rm visible,~\rm \sqrt{s} = 5.02 \rm  ~TeV}(\pt > 2~{\rm GeV}/c, |y| < 0.9)~&=~0.89~\pm~0.15~(\rm stat.) \pm 0.10~(\rm syst.)~{\rm \mu b}.
\nonumber   
\end{align*}

The FONLL calculations, integrated in the corresponding kinematic regions, provide $2.40^{+1.07}_{-0.97}$~${\rm \mu b}$ at $\sqrt{s}$ = 13 TeV and $0.75^{+0.33}_{-0.24}$~${\rm \mu b}$ at $\sqrt{s}$ = 5.02 TeV. The measured visible cross sections are extrapolated down to \pt = 0 relying on the \pt-shape of the FONLL calculations. The extrapolation factors, computed using the same approach as described in Ref.~\cite{Abelev:2012gx}, are  $1.113^{+0.009}_{-0.024}$ and $1.559^{+0.048}_{-0.099}$ at \s = 13 TeV and \s = 5.02 TeV, respectively, which indicates that the measurement at \s = 13 TeV covers about 90\% of the total cross section at midrapidity. The measurement at \s = 5.02 TeV covers only approximately 45\% of the total cross section, half of that covered at 13 TeV, mostly because of the higher value of the minimum  \pt\ limit, 2 GeV/$c$ instead of 1 GeV/$c$.  
 The uncertainties of the extrapolation factors are obtained by changing independently renormalisation and factorisation scales as well as beauty-quark mass and 
 parton distribution functions, considering all variations mentioned above. 
 In addition, a systematic uncertainty  related to the incomplete knowledge of beauty-quark hadronisation fractions was estimated through MC simulations. In particular, the mixture of beauty-flavour hadrons in PYTHIA 6.4 was reweighted in order to match the corresponding measurement of the LHCb collaboration in pp collisions at \s = 13 TeV~\cite{PhysRevD.100.031102}. A systematic uncertainty was computed by comparing the extrapolation factors obtained with and without the application of the reweighting procedure. The corresponding uncertainty of the extrapolation factor is below 1\% at \s = 13 TeV and about 1\% at \s = 5.02 TeV.

The obtained extrapolated \pt-integrated non-prompt \pJPsi\ cross sections per unit of rapidity are:  

\begin{align*}
\frac{{\rm d}\sigma_{\pJPsi \leftarrow {\it h}_{\rm B}}^{\sqrt{s} = 13 \rm TeV}}{{\rm d}y}~&=~1.68~\pm~0.14~(\rm stat.) \pm 0.16 (\rm syst.)^{+0.01}_{-0.04}~(\rm extr.)~{\rm \mu b}, \\
\frac{{\rm d}\sigma_{\pJPsi \leftarrow {\it h}_{\rm B}}^{\sqrt{s} = 5.02 \rm TeV}}{{\rm d}y}~&=~0.77~\pm~0.13~(\rm stat.) \pm 0.09~(\rm syst.)^{+0.02}_{-0.05}~(\rm extr.)~{\rm \mu b}.
\nonumber   
\end{align*}

\begin{figure}[t!]
\centering
  \includegraphics[width=0.49\textwidth]{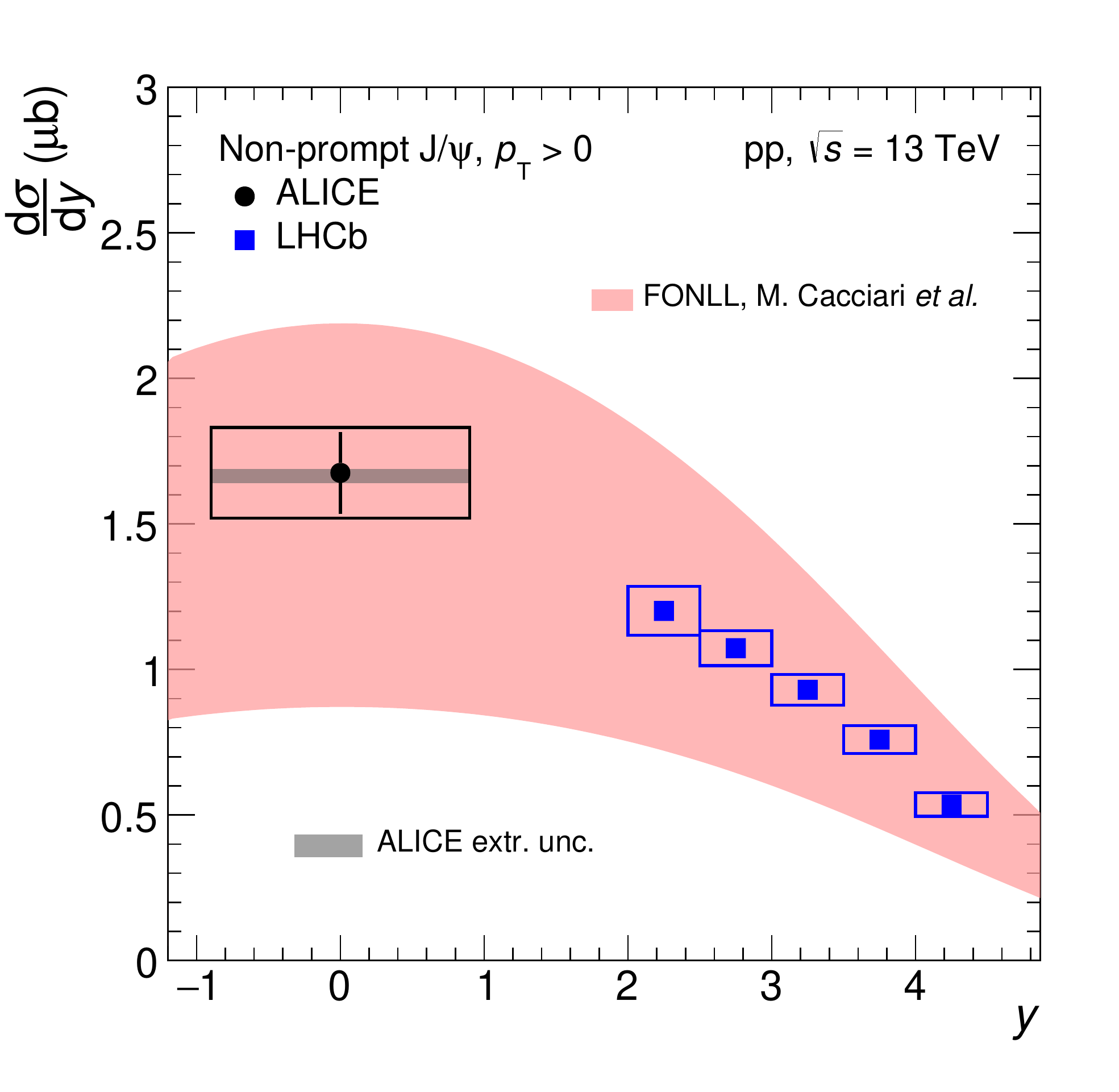}
  \includegraphics[width=0.49\textwidth]{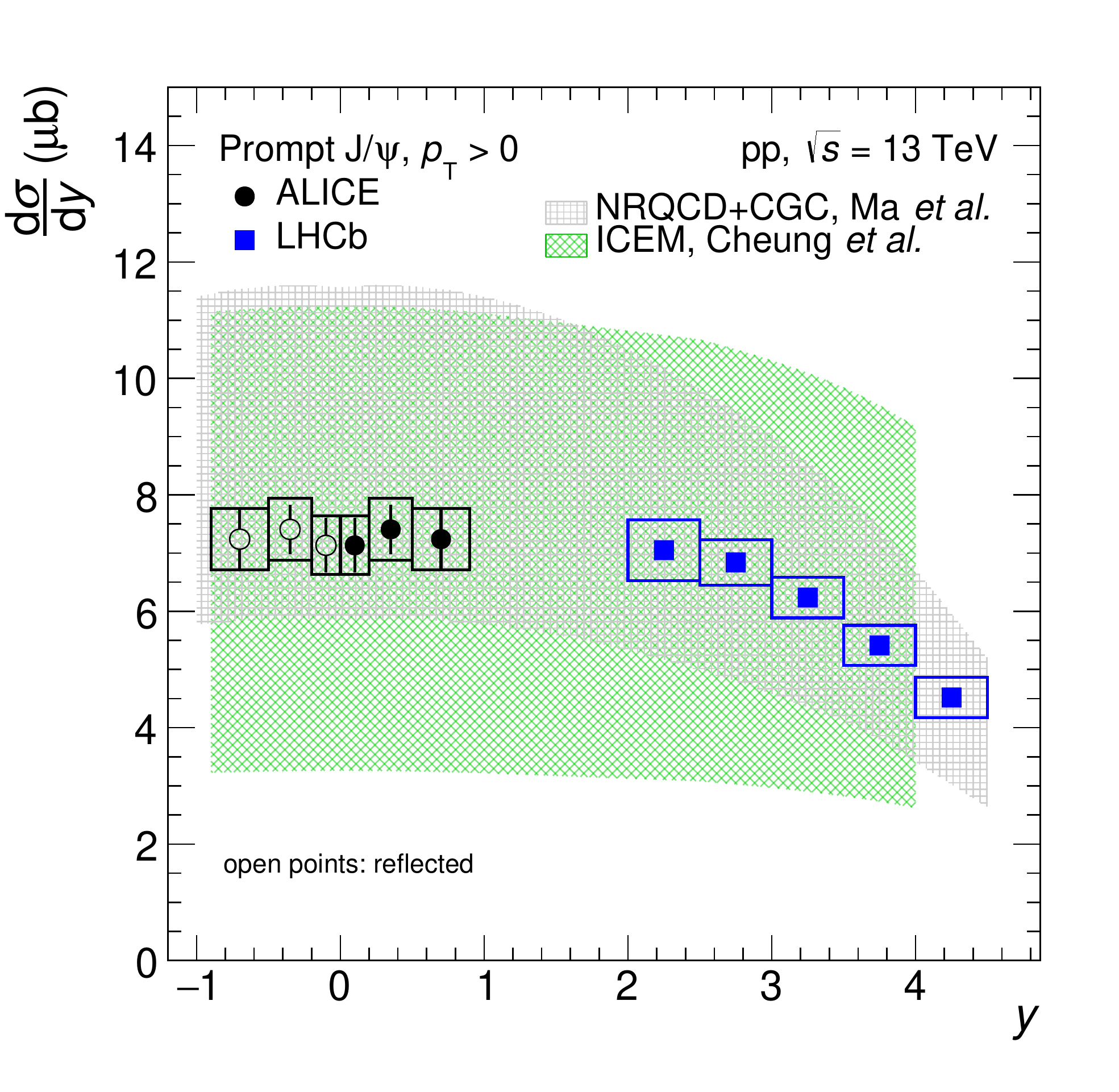}
  \caption{
  Left panel: the $\frac{{\rm d}\sigma}{{\rm d}y}$ of non-prompt \pJPsi extrapolated down to \pt\ = 0 at midrapidity computed by the ALICE collaboration compared with similar measurements in pp collisions at \s = 13 TeV carried out at forward rapidity by the LHCb collaboration~\cite{Aaij:2015rla}.
	Right panel: the $\frac{{\rm d}\sigma}{{\rm d}y}$ of prompt \pJPsi as a function of the rapidity in \pp collisions at \s = 13 TeV. The results close to midrapidity are based on the ALICE measurements extrapolated down to \pt = 0, and closed (open) symbols represent measured (reflected) data points (see text for details). Similar results obtained by the LHCb collaboration~\cite{Aaij:2015rla} at forward rapidity are shown as well. The theoretical calculations are from Refs.~\cite{Cacciari_2001,Cacciari:2012ny,Ma:2014mri,PhysRevD.98.114029}.}
  \label{Fig:dsdydpTNonPromptAndPrompt}
\end{figure}

Although the extrapolation uncertainty is larger at \s = 5.02 TeV than at \s = 13 TeV, it is still negligible compared to the total systematic uncertainty.
The \pt-integrated cross section is compared with similar measurements in pp collisions at \s = 13 TeV performed by LHCb~\cite{Aaij:2015rla} at forward rapidity in the left panel of Fig.~\ref{Fig:dsdydpTNonPromptAndPrompt}. The shadowed area on top of the ALICE point represents the systematic uncertainty which originates from the extrapolation. Theoretical predictions from the FONLL calculations are superimposed on the plot.

The prompt \pJPsi cross section for $\pt > 0$ at midrapidity ($|y| < 0.9$) can be obtained by subtracting the extrapolated non-prompt \pJPsi\ cross section from the inclusive one reported for $\pt > 0$ in Refs.~\cite{Acharya:2019lkw,Alice13TeV}:

\begin{align*}
\frac{{\rm d}\sigma_{\rm prompt~\pJPsi}^{\sqrt{s}~=~13~\rm TeV}}{{\rm d}y}~&=~7.29 \pm 0.27 (\rm stat.) \pm 0.52 (\rm syst.)^{+0.04}_{-0.01} (\rm extr.) {\rm \mu b}, \\
\frac{{\rm d}\sigma_{\rm prompt~\pJPsi}^{\sqrt{s}~=~5.02~\rm TeV}}{{\rm d}y}~&=~4.87 \pm 0.25 (\rm stat.) \pm 0.35 (\rm syst.)^{+0.05}_{-0.02} (\rm extr.) {\rm \mu b}.
\nonumber
\end{align*}
 
The uncertainty from the luminosity is included in the total systematic uncertainty. The \pt-integrated ($\pt > 0$) prompt \pJPsi cross section at 13 TeV was determined additionally in three rapidity intervals (\mbox{$|y| < 0.2$,} \mbox{$0.2 < |y| < 0.5$,} \mbox{$0.5 < |y| < 0.9$}), by combining non-prompt \pJPsi fractions as a function of rapidity discussed in Section~\ref{Sec:dataAnalysis} and inclusive J/$\psi$ cross section measurements in the same rapidity intervals~\cite{Alice13TeV}. The rapidity dependent cross section is shown in the right panel of Fig.~\ref{Fig:dsdydpTNonPromptAndPrompt}, together with the  measurements from LHCb~\cite{Aaij:2015rla} performed at forward rapidity. 
The systematic uncertainties shown in the right panel of Fig.~\ref{Fig:dsdydpTNonPromptAndPrompt}, represented by boxes, include the extrapolation uncertainty as well as the uncertainty from luminosity determination. The measurements are compared with the NRQCD+CGC calculations from Ref.~\cite{Ma:2014mri} and to the ones from the ICEM~\cite{PhysRevD.98.114029} model. 
Although the scale uncertainties lead to rather large uncertainties for their calculations, the two models exhibit a different rapidity dependency, with the data favoring the NRQCD+CGC calculations.

Following a similar approach as the one described in Ref.~\cite{Abelev:2012gx}, the \pt-integrated beauty-quark production cross section per unit of rapidity at midrapidity ($|y_{\rm b}|<$0.9), ${\rm d}\sigma_{\rm b \overline{\rm b}}/{\rm d}y$, can be extracted. The extrapolation was carried out at both centre-of-mass energies, starting from the visible non-prompt \pJPsi\ cross section measurements $\sigma_{\pJPsi \leftarrow {\it h}_{\rm B}}^{\rm visible}$, assuming:

\begin{equation}
{\rm d}\sigma_{\rm b \overline{\rm b}}/{\rm d}y = ({\rm d}\sigma_{\rm b \overline{\rm b}}^{\rm FONLL}/{\rm d}y) \times \frac{\sigma_{\pJPsi \leftarrow {\it h}_{\rm B}}^{\rm visible}}{\sigma_{\pJPsi \leftarrow {\it h}_{\rm B}}^{\rm visible,FONLL}},
\end{equation}

where ${\rm d}\sigma_{\rm b \overline{\rm b}}^{\rm FONLL}/{\rm d}y$ and $\sigma_{\pJPsi \leftarrow {\it h}_{\rm B}}^{\rm visible,FONLL}$ represent the beauty-quark production cross section at midrapidity and the non-prompt \pJPsi\ cross section in the visible region both evaluated using FONLL calculations. The average branching ratio of inclusive beauty-hadrons decaying into \pJPsi\ used for the computation of $\sigma_{\pJPsi \leftarrow {\it h}_{\rm B}}^{\rm visible,~FONLL}$ is $BR(h_{\rm B} \rightarrow \pJPsi + X) = (1.16 \pm 0.10)$\%~\cite{Tanabashi:2018oca}. The resulting beauty-quark production cross sections at midrapidity are thus:

\begin{align*}
	\frac{{\rm d}\sigma_{\rm b \overline{\rm b}}}{{\rm d}{\it y}}_{|{\it y_{\rm b}}|<0.9}^{\sqrt{s} = 13 \rm TeV} &= 73.3 \pm 6.1 (\rm stat.) \pm 9.3 (\rm syst.) _{-2.3}^{+0.8} (\rm extr.)~{\rm \mu b}, \\
	\frac{{\rm d}\sigma_{\rm b \overline{\rm b}}}{{\rm d}{\it y}}_{|{\it y_{\rm b}}|<0.9}^{\sqrt{s} = 5.02 \rm TeV} &= 34.7 \pm 5.9 (\rm stat.) \pm 5.0 (\rm syst.) _{-2.3}^{+1.1} (\rm extr.)~{\rm \mu b}, 
\nonumber     
\end{align*}

where the total systematic uncertainty includes both the uncertainty of the $BR(h_{\rm B} \rightarrow \pJPsi + X)$, which amounts to 8.6\%, and the uncertainty from the luminosity estimation.
The extrapolation uncertainty due to FONLL was computed using the same approach as for the extrapolation of the non-prompt \pJPsi\ cross section down to \pt = 0, including also the systematic uncertainty obtained by changing the beauty-quark hadronisation fractions.
A possible additional uncertainty originating from the assumption of the $BR(h_{\rm B} \rightarrow \pJPsi + X)$ was also investigated. In particular, the world average value~\cite{Tanabashi:2018oca}, used for the computation above, refers to the mixture of beauty-flavour mesons and baryons based on measurements performed at the Large Electron-Positron Collider (LEP). This mixture might be different at the LHC, according to the recent LHCb measurements~\cite{PhysRevD.100.031102}, thus affecting the average $BR(h_{\rm B} \rightarrow \pJPsi + X)$. The extrapolation factor was recomputed through fast simulations considering measurements of the branching ratios of non-strange beauty-flavour mesons decaying into \pJPsi available in the PDG~\cite{Tanabashi:2018oca} and some reasonable variation intervals for those of $\mathrm{B}^{0}_{\rm s}$ and $\Lambda_{\rm b}^{0}$. These intervals were defined by combining the sum of the branching ratios of exclusive decay channels with a \pJPsi in the final state 
available in the PDG~\cite{Tanabashi:2018oca} as well as the beauty-quark hadronisation fractions measured at LEP~\cite{Tanabashi:2018oca}. 
The corresponding uncertainty of the extrapolation factor, obtained by assuming hadronisation fractions from LHCb, was found to be less than 3\%, independent of both the collision energy and the extrapolation region. 
This number is well within the 8.6\% uncertainty of the PDG $BR(h_{\rm B} \rightarrow \pJPsi + X)$, and therefore it is not included as an additional uncertainty of the extrapolation factor.

The ${\rm d } \sigma_{\rm b \overline{\rm b}}/{\rm d}y$  at \s = 5.02 TeV is found to be consistent with the result of a measurement from non-prompt D mesons~\cite{Acharya:2021cqv}, ${\rm d } \sigma_{\rm b \overline{\rm b}}/{\rm d}y_{|y_{\rm b}|<0.9}$~=~32.5~$\pm$~2.3~(stat)~$\pm$~2.5~(syst)$^{+3.8}_{-1.1}(\rm extr.)~{\rm \mu b}$, where the systematic uncertainty includes both uncertainties due to the branching ratio and luminosity. This value was obtained by applying to the published measurement a correction factor of 1.06 evaluated through POWHEG simulations~\cite{Acharya:2021cqv}, in order to convert the rapidity selection criterion of the \bbbar pair ($|y_{\rm b \overline{\rm b}}| < 0.5$) to a rapidity selection criterion on the single beauty-quark ($|y_{\rm b}| < 0.5$). An additional correction factor of about 1\% was needed for obtaining the cross section in the rapidity range 
$|y_{\rm b}| < 0.9$. The weighted average of the two measurements was calculated according to the procedure described in~\cite{LYONS1988110}, assuming the extrapolation uncertainties, estimated through FONLL for both measurements, as well as the systematic uncertainty on the track reconstruction efficiency~\cite{Acharya:2019lkw,Acharya:2021cqv}, to be fully correlated. 
The combined value is   ${\rm d } \sigma_{\rm b \overline{\rm b}}/{\rm d}y_{|y_{\rm b}|<0.9}$~=~32.5~$\pm$~2.2~(stat)~$^{+2.5}_{-2.4}~(\rm syst)^{+3.6}_{-1.1}(\rm extr.)~{\rm \mu b}$, where the systematic uncertainty
includes contributions from the branching ratios and luminosity. 

\begin{figure}[t!]
\centering
  \includegraphics[width=0.61\textwidth]{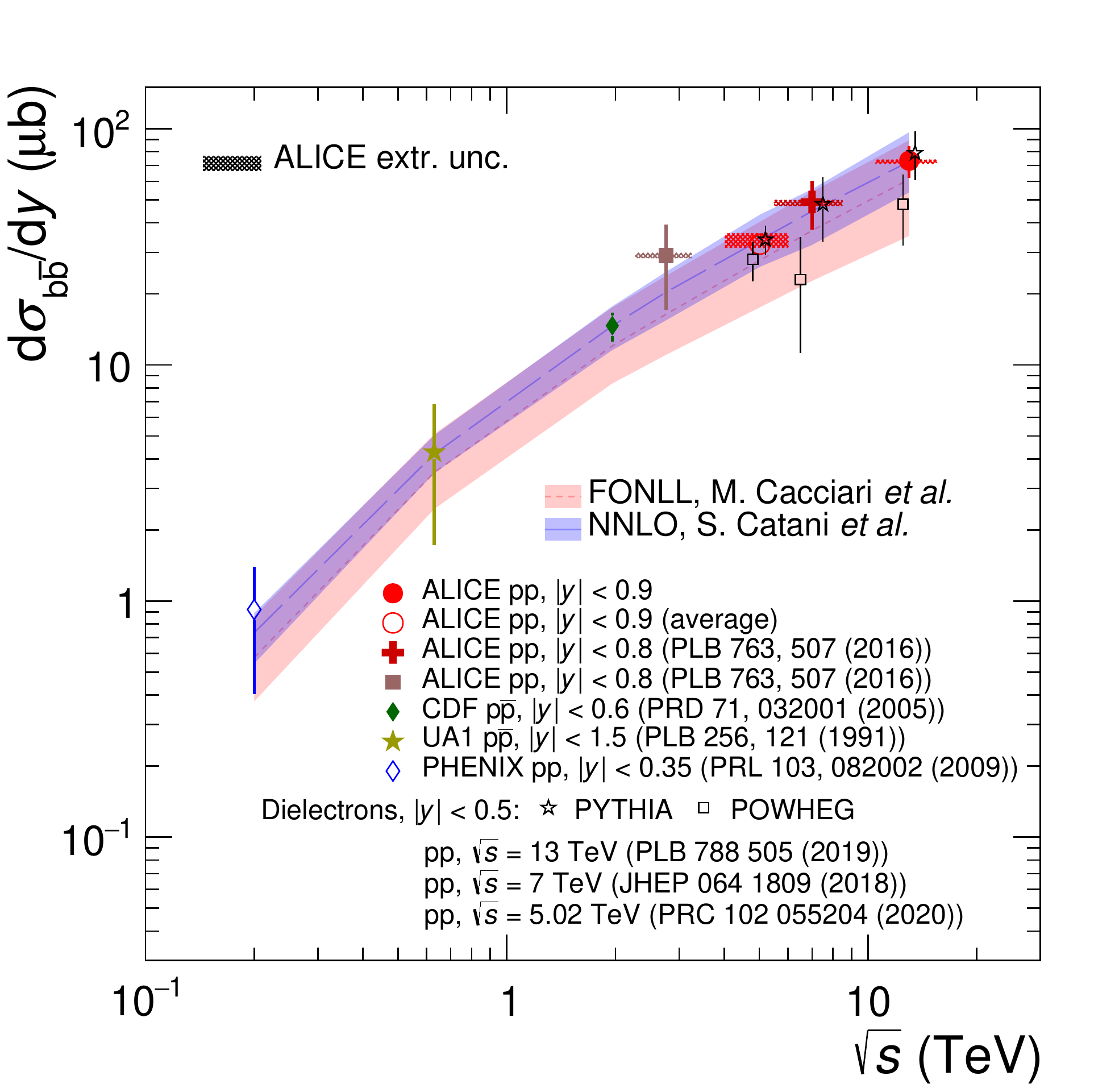}
  \caption{The ${\rm d } \sigma_{\rm b \overline{\rm b}}/{\rm d}y$ at midrapidity as a function of the centre-of-mass energy. The ALICE measurement at \s = 5.02 TeV corresponds to the weighted average of
non-prompt D mesons~\cite{Acharya:2021cqv} and non-prompt \pJPsi (see text for details). The ALICE results are compared with existing measurements in pp collisions (PHENIX~\cite{PhysRevLett.103.082002} and ALICE~\cite{Abelev:2012sca}) and in  p$\overline{\rm p}$ collisions (UA1~\cite{Albajar:213990} and CDF~\cite{PhysRevD.71.032001}). The shaded area around the ALICE data points represents the extrapolation uncertainty. Results from dielectron measurements from the ALICE collaboration, obtained using either PYTHIA or POWHEG simulations, are also shown~\cite{Acharya:2018kkj,Acharya:2018ohw,PhysRevC.102.055204}.
	FONLL~\cite{Cacciari_2001,Cacciari:2012ny} and NNLO~\cite{Catani:2020kkl} calculations in the rapidity range $|y| < 0.9$, with the corresponding uncertainty bands, are superimposed.  }
  \label{Fig:BeautyCrossSec}
\end{figure}

The ${\rm d } \sigma_{\rm b \overline{\rm b}}/{\rm d}y$ computed at midrapidity in pp collisions at \s = 5.02 and 13 TeV are shown as a function of centre-of-mass energy in Fig.~\ref{Fig:BeautyCrossSec}, together with existing experimental measurements in pp collisions from PHENIX~\cite{PhysRevLett.103.082002} and ALICE~\cite{Abelev:2012sca}, and results in p$\overline{\rm p}$ collisions from UA1~\cite{Albajar:213990} and CDF~\cite{PhysRevD.71.032001}. Beauty-quark production cross sections from ALICE dielectron measurements, extrapolated using either PYTHIA or POWHEG simulations~\cite{Acharya:2018kkj,Acharya:2018ohw,PhysRevC.102.055204}, are shown as well. The depicted FONLL calculations are in agreement with the data, although the experimental points sit on the upper side of the theoretical uncertainties. 
The \pt-integrated \bbbar cross sections in Fig.~\ref{Fig:BeautyCrossSec} are also compared with calculations with next-to-next-to-leading-order (NNLO) QCD radiative
corrections~\cite{Catani:2020kkl}, which recently became available. The NNLO calculations are found to be slightly higher than FONLL, resulting in a general better 
description of the measurements. 

The total \bbbar production cross section, obtained by extrapolating in rapidity and down to \pt = 0, is

\begin{equation}
\sigma({\rm pp} \rightarrow {\rm b}\overline{\rm b} + X) = \alpha_{4\pi} \times \frac{\sigma_{\pJPsi \leftarrow {\it h}_{\rm B}}^{\rm visible}}{2 \times BR(h_{\rm B} \rightarrow \pJPsi + X) },
\end{equation}  

where the extrapolation factor, $\alpha_{4\pi}$, is the ratio of the \bbbar cross section in the full phase space to the visible non-prompt \pJPsi\ cross section. The factor 2 in the denominator takes into account that beauty quarks are produced in pairs and the non-prompt \pJPsi can originate from the decay of hadrons containing either a b or a $\overline{\rm b}$ quark. The extrapolation factor is computed using the FONLL calculation and found to be $\alpha_{4\pi}^{\s~=~13~{\rm TeV}} = 4.63^{+0.09}_{-0.11}$ at 13 TeV and $\alpha_{4\pi}^{\s~=~5.02~{\rm TeV}} = 5.69^{+0.21}_{-0.24}$ at 5.02 TeV. The extrapolation uncertainties are evaluated using the same approach as described for the extrapolation of the ${\rm d } \sigma_{\rm b \overline{\rm b}}/{\rm d}y$ at midrapidity. The corresponding total \bbbar cross sections are: 

\begin{align*}
    \sigma_{\rm b \overline{\rm b}}^{\sqrt{s} = 13 \rm TeV}~&=~541 \pm 45 (\rm stat.) \pm 69 (\rm syst.)_{-12}^{+10} (\rm extr.)~{\rm \mu b}, \\
    \sigma_{\rm b \overline{\rm b}}^{\sqrt{s} = 5.02 \rm TeV}~&=~218 \pm 37 (\rm stat.) \pm 31 (\rm syst.)_{-9.1}^{+8.2} (\rm extr.)~{\rm \mu b}.
\end{align*}

The total systematic uncertainties include contributions from $BR(h_{\rm B} \rightarrow \pJPsi + X)$ and the luminosity. The measured values can be compared with the predictions from FONLL (NNLO), namely $\sigma_{\rm b \overline{\rm b}, {\rm FONLL}}^{\sqrt{s} = 13 \rm TeV} = 472^{+219}_{-190}~{\rm \mu b}$ ($\sigma_{\rm b \overline{\rm b}, {\rm NNLO}}^{\sqrt{s} = 13 \rm TeV} = 508^{+168}_{-132}~{\rm \mu b}$) and $\sigma_{\rm b \overline{\rm b},{\rm FONLL}}^{\sqrt{s} = 5.02 \rm TeV} = 184^{+85}_{-65}~{\rm \mu b}$ ($\sigma_{\rm b \overline{\rm b},{\rm NNLO}}^{\sqrt{s} = 5.02 \rm TeV} = 206^{+58}_{-47}~{\rm \mu b}$). The experimental results are larger than the central values from both FONLL and NNLO, but in agreement within the large theoretical uncertainties at both centre-of-mass energies. The value at \s = 13 TeV is compatible within uncertainties with the measurement from the LHCb collaboration, based on the non-prompt \pJPsi cross section measured at forward rapidity ($2.0 < y < 4.5$) for \pt $>$ 0, $\sigma_{\rm b \overline{\rm b}} = 495 \pm 2 (\rm stat.) \pm 52 (\rm syst.)~{\rm \mu b}$~\cite{Aaij:2015rla}. The systematic uncertainty quoted by the LHCb collaboration does not include any extrapolation uncertainty, and the extrapolation factor $\alpha_{4\pi}$ = 5.2 was evaluated through PYTHIA6 simulations. The corresponding values obtained with PYTHIA8 simulations and FONLL calculations, both quoted in Ref.~\cite{Aaij:2015rla}, amount to 5.1 and 5.0, respectively.

The combination of the ALICE and LHCb measurements of non-prompt \pJPsi in the respective visible region allows us the determination of the total \bbbar cross section at \s = 13 TeV with the smallest extrapolation factor. 
It is again computed with FONLL calculations as the ratio between the \bbbar cross section in the full phase space over the non-prompt \pJPsi cross section in the combination of the visible regions of ALICE and LHCb. Its value is $\alpha_{\rm 4\pi,ALICE+LHCb}^{\rm FONLL} = 2.395 \pm 0.014$ and it further reduces to $\alpha_{\rm 4\pi,ALICE+LHCb}^{\rm FONLL} = 1.616^{+0.007}_{-0.010}$ by reflecting the LHCb data points around $y = 0$. The resulting total beauty-quark production cross section using the combined measurements is

\begin{equation}
\nonumber
\sigma_{\rm b \overline{\rm b}}^{\sqrt{s} = 13 \rm TeV} = 502 \pm 16 (\rm stat.) \pm 51 (\rm syst.)_{-3}^{+2} (extr.)~{\rm \mu b}, 
\end{equation}

where the uncertainty was evaluated assuming the uncertainties of the ALICE and LHCb non-prompt \pJPsi cross section measurements to be fully uncorrelated.

\section{Summary}
\label{Sec:summary}

The \pt-differential cross sections of prompt and non-prompt \pJPsi\ were measured in the rapidity range $|y| < 0.9$ in pp collisions at \s = 5.02 and 13 TeV down to \pt = 2 GeV/$c$ and 1 GeV/$c$, respectively. In addition, the prompt  \pJPsi\ cross section was measured in three rapidity intervals at \s = 13 TeV. The measured cross sections, both \pt~(or $y$)-differential and integrated over \pt\ 
and $y$, were obtained assuming prompt \pJPsi\ to be unpolarised. Results were compared with theoretical calculations from QCD based models and similar measurements from other LHC experiments. The non-prompt \pJPsi\ cross sections are described by predictions from FONLL calculations at both energies. The prompt \pJPsi\ cross sections as a function of transverse momentum and rapidity are described within uncertainties by models based on NRQCD calculations as well as with an improved version of the CEM. 
The large uncertainties of the model calculations, which arise from the charm quark mass, as well as factorisation and renormalisation scales, do not allow to discriminate among different models. 
The \pt-differential cross sections, for both prompt and non-prompt \pJPsi\, at \s = 5.02 TeV are consistent with complementary measurements from the ATLAS and CMS collaborations, available at high transverse momentum. The ${\rm d}\sigma_{\rm b \overline{b}}/{\rm d}y$ at midrapidity and the total \bbbar cross section were derived by using FONLL \pt\ and $y$ shapes at both centre-of-mass energies. The total \bbbar cross section at \s = 13 TeV is found to be consistent with the measurement from the LHCb collaboration, and a value obtained from the combination of ALICE and LHCb measurements with a significantly reduced extrapolation factor was also provided. 


\newenvironment{acknowledgement}{\relax}{\relax}
\begin{acknowledgement}
\section*{Acknowledgements}

The ALICE Collaboration would like to thank all its engineers and technicians for their invaluable contributions to the construction of the experiment and the CERN accelerator teams for the outstanding performance of the LHC complex.
The ALICE Collaboration gratefully acknowledges the resources and support provided by all Grid centres and the Worldwide LHC Computing Grid (WLCG) collaboration.
The ALICE Collaboration acknowledges the following funding agencies for their support in building and running the ALICE detector:
A. I. Alikhanyan National Science Laboratory (Yerevan Physics Institute) Foundation (ANSL), State Committee of Science and World Federation of Scientists (WFS), Armenia;
Austrian Academy of Sciences, Austrian Science Fund (FWF): [M 2467-N36] and Nationalstiftung f\"{u}r Forschung, Technologie und Entwicklung, Austria;
Ministry of Communications and High Technologies, National Nuclear Research Center, Azerbaijan;
Conselho Nacional de Desenvolvimento Cient\'{\i}fico e Tecnol\'{o}gico (CNPq), Financiadora de Estudos e Projetos (Finep), Funda\c{c}\~{a}o de Amparo \`{a} Pesquisa do Estado de S\~{a}o Paulo (FAPESP) and Universidade Federal do Rio Grande do Sul (UFRGS), Brazil;
Ministry of Education of China (MOEC) , Ministry of Science \& Technology of China (MSTC) and National Natural Science Foundation of China (NSFC), China;
Ministry of Science and Education and Croatian Science Foundation, Croatia;
Centro de Aplicaciones Tecnol\'{o}gicas y Desarrollo Nuclear (CEADEN), Cubaenerg\'{\i}a, Cuba;
Ministry of Education, Youth and Sports of the Czech Republic, Czech Republic;
The Danish Council for Independent Research | Natural Sciences, the VILLUM FONDEN and Danish National Research Foundation (DNRF), Denmark;
Helsinki Institute of Physics (HIP), Finland;
Commissariat \`{a} l'Energie Atomique (CEA) and Institut National de Physique Nucl\'{e}aire et de Physique des Particules (IN2P3) and Centre National de la Recherche Scientifique (CNRS), France;
Bundesministerium f\"{u}r Bildung und Forschung (BMBF) and GSI Helmholtzzentrum f\"{u}r Schwerionenforschung GmbH, Germany;
General Secretariat for Research and Technology, Ministry of Education, Research and Religions, Greece;
National Research, Development and Innovation Office, Hungary;
Department of Atomic Energy Government of India (DAE), Department of Science and Technology, Government of India (DST), University Grants Commission, Government of India (UGC) and Council of Scientific and Industrial Research (CSIR), India;
Indonesian Institute of Science, Indonesia;
Istituto Nazionale di Fisica Nucleare (INFN), Italy;
Institute for Innovative Science and Technology , Nagasaki Institute of Applied Science (IIST), Japanese Ministry of Education, Culture, Sports, Science and Technology (MEXT) and Japan Society for the Promotion of Science (JSPS) KAKENHI, Japan;
Consejo Nacional de Ciencia (CONACYT) y Tecnolog\'{i}a, through Fondo de Cooperaci\'{o}n Internacional en Ciencia y Tecnolog\'{i}a (FONCICYT) and Direcci\'{o}n General de Asuntos del Personal Academico (DGAPA), Mexico;
Nederlandse Organisatie voor Wetenschappelijk Onderzoek (NWO), Netherlands;
The Research Council of Norway, Norway;
Commission on Science and Technology for Sustainable Development in the South (COMSATS), Pakistan;
Pontificia Universidad Cat\'{o}lica del Per\'{u}, Peru;
Ministry of Education and Science, National Science Centre and WUT ID-UB, Poland;
Korea Institute of Science and Technology Information and National Research Foundation of Korea (NRF), Republic of Korea;
Ministry of Education and Scientific Research, Institute of Atomic Physics and Ministry of Research and Innovation and Institute of Atomic Physics, Romania;
Joint Institute for Nuclear Research (JINR), Ministry of Education and Science of the Russian Federation, National Research Centre Kurchatov Institute, Russian Science Foundation and Russian Foundation for Basic Research, Russia;
Ministry of Education, Science, Research and Sport of the Slovak Republic, Slovakia;
National Research Foundation of South Africa, South Africa;
Swedish Research Council (VR) and Knut \& Alice Wallenberg Foundation (KAW), Sweden;
European Organization for Nuclear Research, Switzerland;
Suranaree University of Technology (SUT), National Science and Technology Development Agency (NSDTA) and Office of the Higher Education Commission under NRU project of Thailand, Thailand;
Turkish Energy, Nuclear and Mineral Research Agency (TENMAK), Turkey;
National Academy of  Sciences of Ukraine, Ukraine;
Science and Technology Facilities Council (STFC), United Kingdom;
National Science Foundation of the United States of America (NSF) and United States Department of Energy, Office of Nuclear Physics (DOE NP), United States of America.
\end{acknowledgement}

\bibliographystyle{utphys}   
\bibliography{bibliography}

\newpage
\appendix

%
%

\section{The ALICE Collaboration}
\label{app:collab}

\begin{flushleft} 

\bigskip 

S.~Acharya$^{\rm 143}$, 
D.~Adamov\'{a}$^{\rm 98}$, 
A.~Adler$^{\rm 76}$, 
G.~Aglieri Rinella$^{\rm 35}$, 
M.~Agnello$^{\rm 31}$, 
N.~Agrawal$^{\rm 55}$, 
Z.~Ahammed$^{\rm 143}$, 
S.~Ahmad$^{\rm 16}$, 
S.U.~Ahn$^{\rm 78}$, 
I.~Ahuja$^{\rm 39}$
% ...

\bigskip

\bigskip 

\textbf{\Large Affiliation Notes}

\bigskip 

% ...
$^{\rm VII}$ Also at: University of Kansas, Lawrence, Kansas, United States\\

\bigskip

\bigskip 

\textbf{\Large Collaboration Institutes}

\bigskip 

% ...
$^{16}$ Department of Physics, Aligarh Muslim University, Aligarh, India\\
% ...
$^{31}$ Dipartimento DISAT del Politecnico and Sezione INFN, Turin, Italy\\
% ...
$^{35}$ European Organization for Nuclear Research (CERN), Geneva, Switzerland\\
% ...
$^{39}$ Faculty of Science, P.J.~\v{S}af\'{a}rik University, Ko\v{s}ice, Slovakia\\
% ...
$^{55}$ INFN, Sezione di Bologna, Bologna, Italy\\
% ...
$^{76}$ Johann-Wolfgang-Goethe Universit\"{a}t Frankfurt Institut f\"{u}r Informatik, Fachbereich Informatik und Mathematik, Frankfurt, Germany\\
% ...
$^{78}$ Korea Institute of Science and Technology Information, Daejeon, Republic of Korea\\
% ...
$^{98}$ Nuclear Physics Institute of the Czech Academy of Sciences, \v{R}e\v{z} u Prahy, Czech Republic\\
% ...
$^{143}$ Variable Energy Cyclotron Centre, Homi Bhabha National Institute, Kolkata, India\\
% ...

\bigskip 

\end{flushleft} 
  
\end{document}